\catcode`\@=11
{\count255=\time\divide\count255 by 60
\xdef\hourmin{\number\count255}
        \multiply\count255 by-60\advance\count255 by\time
   \xdef\hourmin{\hourmin:\ifnum\count255<10 0\fi\the\count255}}
\def\ps@draft{\let\@mkboth\@gobbletwo
    \def\@oddhead{}
    \def\@oddfoot
       {\hbox to 7 cm{$\scriptstyle Draft\ version:\ \draftdate$
       \hfil}
       \hskip -7cm\hfil\rm\thepage \hfil}
    \def\@evenhead{}\let\@evenfoot\@oddfoot}
\catcode`\@=12
\def\draftdate{\number\month/\number\day/\number\year\ \ \ \hourmin }
\def\draft{\pagestyle{draft}\thispagestyle{draft}}
\documentstyle[12pt]{article}
\newcommand{\BE}{\begin{eqnarray}}
\newcommand{\EN}{\end{eqnarray}}
\newcommand{\be}{\begin{equation}}
\newcommand{\en}{\end{equation}}
\newcommand{\non}{\nonumber}
\newcommand{\no}{\noindent}
\newcommand{\vs}{\vspace}

\newcommand{\p}{\partial}
\newcommand{\ha}{{1\over 2}}

\begin{document}
\title{Stability of Moving Fronts in the Ginzburg-Landau Equation }

\author{J.Bricmont\thanks{Supported by EC grant SC1-CT91-0695}
\\UCL, Physique Theorique,
 Louvain-la-Neuve, Belgium\and
A.Kupiainen\thanks{Supported by NSF grant DMS-8903041}
\\Rutgers University, Mathematics Department,\\ New Brunswick
NJ 08903, USA}

\date{}

\maketitle
\begin{abstract}
We use Renormalization Group ideas to study stability
of moving fronts in the Ginzburg-Landau equation in
one spatial dimension. In particular, we prove stability
of the  real fronts under complex perturbations. This
extends the results of Aronson and Weinberger to situations
where the maximum principle is inapplicable and constitutes
a step in proving the general marginal stability hypothesis
for the Ginzburg-Landau equation.

\end{abstract}
\section{Introduction}
\draft

There are very few general approaches to the study of long time existence
and
asymptotics of solutions of nonlinear parabolic partial differential
equations. Typically one has to resort to the use of positivity
properties of the linear semigroup, e.g. the use of maximum
principle and then use  comparison theorems together with compactness
arguments to obtain the asymptotics. Such approaches usually
work only for equations of special form, e.g. with second order
linear part, and positive initial data.

It was noted in \cite{B,G1,G2}, that scaling and renormalization
group (RG) concepts that were very successfull in statistical mechanics
and quantum field theory are also applicable to this study. In
\cite{BKL,BK} we have been developing a mathematical RG theory to
prove global existence and detailed long time asymptotics for
classes of nonlinear parabolic equations.
The RG approach does not depend on the applicability of positivity
conditions such as the maximum principle. The theory
moreover shows
how {\it universality} emerges in such equations: the long time
asymptotics is independent on the initial data and the equation
within classes of data and equations.

In this paper we study
one such problem lacking positivity,
the stability of moving front solutions in the
Ginzburg-Landau equation
\be
\dot{u}=\p^2 u+u-|u|^2u
\en
where $u:{\bf R}\times{\bf R}\rightarrow{\bf C}$, is
complex, $\p=
{\p\over\p x}$ and dot denotes the time derivative.

There is an extensive study \cite{AW,Br}
of the stability of moving front solutions
to (1):
\be
u=r_c (x-ct)
\en
where $r_c$ is a real (and non-negative) function. The analysis is
based on the use of the maximum principle and applies to the case of
positive initial data of (1). However, for complex data very little
is known, becuse the maximum principle is no longer applicable. We
show in this paper how the RG ideas can be used to prove the
stability of the real front solutions (2) under complex perturbations.

Equation (1) has been an important model in the physics literature
for the study of velocity selection for propagating patterns
\cite{DL,D,B-J}. (1) has solutions (2) for a whole range of values
of $c$ (see Section 2) and, for $c\geq 2$ the solutions are
linearly stable in a space of functions with a prescribed
$c$-dependent exponential decay at infinity. A natural
question is: which of these moving fronts, if any, is
selected for an initial data of compact support (this is
also the physically realistic case)? The marginal stability
hyphothesis \cite{DL} states that the marginally stable
$c=2$ is the one selected. For real positive initial data
this indeed is proven by Aronson and Weinberger \cite{AW,Br}.
For complex data there is a large set of complex front
solutions \cite{B-J,CE1}, many of which are linearly
stable \cite{CE1}. An outstanding problem is to prove or
disprove the marginal stability hypothesis for complex
compactly supported data (for an argument, see \cite{B-J}).
While this is still beyond our methods, the present paper
can be considered as a step in this direction.

\section{Results}

\setcounter{equation}{0}

We write the Ginzburg-Landau equation in radial and angle variables,
${\bf u}= re^{i\varphi}$:
\be
\dot r = \partial^2 r + r (1 -(\partial \varphi)^2) - r^3
\en
\be
\dot \varphi = \partial^2 \varphi + 2 r^{-1} \partial \varphi \partial r
\en

It is well known that these equations have real, positive, front solutions,
i.e.
solutions of the form
\be
\varphi=0, r=r_c (x-ct) \geq 0,
\en
such that $r_c$ interpolates between a stable and an unstable solution of
(1), i.e.
$r_c \rightarrow +1$ for $x \rightarrow - \infty, r_c \rightarrow 0$, for
$x
\rightarrow + \infty$. Indeed, from (1), we see that $r_c$ satisfies
\be
r_c^{''} + c r'_c + r_c - r^3_c = 0
\en
which, if we reinterpret the variable as ``time", can be seen as Newton's
equation of
motion of a particle of mass one subjected to a friction term $c r'_c$ and
to a force
deriving from the potential $\frac{r^2}{2} - \frac{r^4}{4}$, which is an
inverted
double-well. It is intuitively clear and easily proved that, for $c$ not
too small,
solutions exists that satisfy the required conditions, i.e. such that $r_c$
tends, as
``time" goes to $+ \infty$, to zero, the stable critical point of the
potential, and
to one as ``time" goes to $- \infty$. For large ``time" $u,r_c (u)$ will
decay
exponentially, as is seen from the linearization of (4) at $r=0$. One
gets
\be
r_c (u) \leq (c_1+c_2u) e^{- \gamma u}
\en
where $\gamma$ is given by
$\gamma^2 - c\gamma + 1 = 0 $ i.e.
\be
\gamma_c = \ha(c - \sqrt{c^2 -4})
\en
which is real for $c \geq 2$, in which case $ \gamma_c \leq 1$
(actually, one can take $c_2=0$ in (5), if $\gamma <1$).
Thus, the larger the friction,
the slower the decay. For $c <2$,
the
solution ``overshoots" the minimum at zero, i.e. $r_c$ is no longer
positive. Each of
the solutions $r_{c}$ with $c \geq 2$ is stable under real perturbations
$(\varphi =
0):$  if we start with initial data $r(x,0)$, with $r=r_{c} + s $ with
$0 \leq
r \leq 1$, $s$ decaying faster than $e^{- \gamma_c x}$ for $x \rightarrow +
\infty$,
$r(x,t)$ will converge, as $t \rightarrow + \infty$, to $r_{c} (x-ct)$,
see \cite{AW,Br}.

However the solution with $c=2, \gamma_c =1$ is more stable than the others
in the
sense that any initial data $r (x,0)$ with $0 \leq r \leq 1$ which decays
faster than
$e^{-x}$ as $x \rightarrow + \infty$ (in particular, if $r$ is of compact
support)
will converge, as $t \rightarrow + \infty$, to $r_{2} (x-2t)$
\cite{AW,Br}.

Now we consider a complex perturbation of $r_{c} : r(x,0) = r_{c} + s$
with $\varphi
(x,0) \neq 0$ and $s(x,0)$ small in a suitable sense. The equations
satisfied by
$\varphi$ and $s$ are :
\be
\dot \varphi = \partial^2 \varphi + 2(\partial r_{c}\partial
\varphi + \partial s \partial \varphi) (r_{0,c} + s)^{-1}
\en
\be
\dot s = \partial^2 s + s (1-3 r^2_{c})
- r_{c} (\partial \varphi)^2
- s (\partial \varphi)^2 -3 r_{c} s^2 - s^3
\en

We want to solve these equations for all $t\geq 1$ with a given
initial data $\varphi(x,1)=\varphi(x), s(x,1)=s(x)$ (it will be
notationally convenient to take the initial time $t=1$). We will
state the main result only in the hardest case $c=2$ and will
comment on $c>2$ in the course of the proof. We consider the
initial data in the Banach space of $C^1$-functions $\varphi,s$ with
the norm
\be
\|(\varphi,s)\|=\sup_x(1+\mid x \mid)^{3+\delta}(
\mid \varphi (x) \mid + \mid  \varphi' (x) \mid+(1+e^x)(
\mid s (x) \mid + \mid  s' (x) \mid))
\en
and prove the

\vs{3mm}

\no {\bf Theorem}.{\it
For any $\delta > 0$ there exists an $\epsilon > 0$ such that equations
$(7,8)$ with $c=2$ and
initial data $\varphi (x,1) = \varphi (x), s(x,1) =
s(x)$ with $\|(\varphi,s)\|<\epsilon$
have a unique $C^1$ solution $\varphi (x,t),
s(x,t)$, for
all $t \geq 1$, such that}
\be
\| \varphi (\cdot,t) \|_{\infty}, \|s(\cdot,t) \|_{\infty}
\leq t^{- \frac{1}{2} + \delta}
\en

\no{\bf Remark}.
In the course of the proof, we shall give more detailed and slightly
sharper
bounds and we shall also prove bounds on
$\partial \varphi (x,t), \partial s (x,t).$

\vs{5mm}

Before we go to the proof of the Theorem, we want to discuss the
result in an informal manner.
Since $r_{c}$ is a function of $x-ct$ it is convenient to consider also
the
equation in the frame of reference of the front :
let $u=x-ct$
and $\varphi_f (u,t) = \varphi (u+ct,t)$,$s_f (u,t) = s (u+ct,t)$; then
$\varphi_f$ and $s_f$ satisfy equations like (7,8), with $c \p\varphi_f$
added
to the RHS of (7) and $c \p s_f$ added to the one of (8). Now $r_c = r_c
(u)$
is time-independent.

To understand the expected behaviour of $\varphi_f (u,t)$, let us consider
the
linearised equation around the zero solution :
\be
\dot \varphi_f = \partial^2 \varphi_f + 2 \partial r_{c} \partial
\varphi_f r_c^{-1} + c
\partial \varphi_f
\en

It is convenient to rewrite this equation as an imaginary
time Schr\"odinger equation: Let
\be
\varphi_f (u,t) = e^{- \frac{c}{2}u}r_{c}(u)^{-1} \psi (u,t)
\en
Then $\psi$ satisfies
\be
\dot \psi = \partial^2 \psi - V\psi
\en
with
\be
V = \frac{c^2}{4} + \frac{r''_{c}}{r_{c}} + c \frac{r'_{c}}{r_{c}}
=
\frac{c^2}{4} - 1 + r^2_{c}.
\en

To derive the last equality, we used eq.(4), satisfied by $r_c$. Since
$r_{c} \simeq
1$ for $u \rightarrow - \infty$, $r_{c} \simeq e^{- \gamma u}$ for $u
\rightarrow
+ \infty$, we have $V \simeq c^2/4$ for $u \rightarrow - \infty, V \simeq
c^2/4-1$ for
$u \rightarrow + \infty$. So, starting with $\psi (u,0)$ localised around
$u=0$, we
expect
\be
\psi (u,t) \sim
\left \{
\begin{array}{ll}
\frac{e^{- \frac{c^2t}{4}-\frac{u^2}{4t}}}{\sqrt t} & u \rightarrow -
\infty \\
\frac{e^{- (\frac{c^2}{4}-1)t-\frac{u^2}{4t}}}{\sqrt t} & u \rightarrow +
\infty
\end{array}
\right.
\en

Hence,
\be
\varphi_f (u,t) \sim
\left \{
\begin{array}{ll}
\frac{e^{- \frac{c^2t}{4}-\frac{u^2}{4t}- \frac{c}{2}u}}{\sqrt t} & u
\rightarrow -
\infty \\
\frac{e^{- (\frac{c^2}{4}-1)t-\frac{u^2}{4t}-
(\frac{c}{2}-\gamma)u}}{\sqrt t} & u
\rightarrow + \infty
\end{array}
\right.
\en
which, using (6) under the form $\frac{c^2}{4}-1 =
\frac{(c-2\gamma)^2}{4}$, can be
written as
\be
\varphi_f (u,t) \sim
\left \{
\begin{array}{ll}
\frac{e^{- \frac{(u + ct)^2}{4t}}}{\sqrt t} & u \rightarrow - \infty \\
\frac{e^{- \frac{(u+(c-2\gamma)t)^2}{4t}}}{\sqrt t} & u \rightarrow +
\infty
\end{array}
\right.
\en

Since $u+ct=x$, the first part of (17) is a diffusive wave stationary in
the fixed
frame. Since $c-2 \gamma \geq 0$ (see (6)), the second part vanishes
rapidly, as
$ u \rightarrow + \infty$, except when $c=2, \gamma = 1$. There $\varphi$
contains also a diffusive wave which is
``carried along" by the front. This is a rough, but basically correct
picture.
However, we shall see later that a proper treatment of the effect of the
potential
yields a different power law than $\frac{1}{\sqrt t}$ in (17). From now on,
we shall
concentrate on the most interesting front namely the one with $c=2, \gamma
= 1$ and
write $r$ for $r_{2}$ .

Let us consider the linear equation for $s_f$ in that case, in the front
frame :
\be
\dot s_f = \partial^2 s_f + 2 \partial s_f + s_f (1-3 r^2)
\en
Writing $s_f=e^{-u} \sigma$, we get
\be
\dot \sigma = \partial^2 \sigma - \tilde{V} \sigma
\en
with $\tilde{V} = 3 r^2$.

Following the analysis leading to (17), we get
\be
s(u,t) \sim
\left \{
\begin{array}{ll}
\frac{
e^{-3t-\frac{u^2}{4t}-u}
}
{\sqrt t} =
\frac{
e^{-2t}}{\sqrt t}
e^{-\frac{
(u+2t)^2}{4t}} & \mbox{as} \;\;\;  u \rightarrow - \infty \\
\frac{
e^{-\frac{u^2}{4t}}}
{\sqrt t} e^{-u} & \mbox{as} \;\;\;  u
\rightarrow + \infty
\end{array}
\right.
\en

There is a``wave" which is stationary in the front frame, but exponentially
decreasing in $u$ , while the wave which stays in the fixed frame is
suppressed by the factor $e^{-2t}$.
Of course, when we take into account the nonlinear terms in (7,8), we see
that a term
like $- r (\partial \varphi)^2$ will prevent $s$ from decaying
exponentially in time,
due to the slower decay of $ \varphi$, see (17).

\section{The Proof}
\setcounter{equation}{0}
\vs{5mm}

\no{\bf 3.1. Preliminaries: the RG-setup}

\vs{3mm}

The proof of the Theorem is based on the
Renormalization Group method
\cite{G1,BKL,BK}, which for the purposes of the
present work consists of the Picard iteration
composed with a scaling. Let us define
\BE
\varphi_n (x,t) & \equiv & \varphi (L^n x, L^{2n}t) \\
s_n (x,t) & \equiv & s (L^n x, L^{2n}t)
\EN
and
\BE
\Phi_n (x) &\equiv& \varphi (L^n x, L^{2n}) =\varphi_n (x,1) \\
S_n (x) &\equiv& s (L^n x, L^{2n})= s_n (x,1)
\EN

We shall prove, inductively in $n$, bounds on
$\varphi_n , s_n$ for the finite time interval
[1,$L^2$], which will imply (2.10). This is done
by controlling the "RG-map" ${\cal R}_n$, which relates
the $\Phi_n,S_n$ to $\Phi_{n+1},S_{n+1}$:
\be
{\cal R}_n (\Phi_n,S_n)  =(\Phi_{n+1} ,S_{n+1})
\en
${\cal R}_n$ is constructed by solving a scaled version of
the original equation (2.7,2.8) for a finite time.
For this, it will be more convenient to work in the variable $u$
relative to the moving frame  of the front,
which under the scaling (1),(2) becomes
\be
u \equiv x-2 L^n t
\en
($c=2$ here) so that, if we define
\BE
\varphi_{fn}(u,t) &\equiv& \varphi_n (u+2 L^n t, t) \\
s_{fn} (u,t) &\equiv& s_n (u+2 L^n t, t)
\EN
we have
\BE
\varphi_{fn} (u,t) &=& \varphi_f (L^n u, L^{2n}t) \\
s_{fn} (u,t) &=& s_f (L^n u, L^{2n}t).
\EN
We will from now on drop the subscript $f$, and use the
variables $x$ and $u$ to distinguish the frame we are in.
$\varphi_n (u,t), s_n (u,t)$ solve the equations
\BE
\dot{\varphi}_n & = &\p^2\varphi_n +
2 L^n \p \varphi_n + 2 L^n \p\varphi_n
\left(
\frac{q_n + L^{-n} \p s_n / r_n}{1 + s_n / r_n}
\right) \\
\dot{s}_n & = & \p^2 s_n + 2 L^n \p s_n
+ L^{2n} s_n (1-3 r^2_n) - r_n(\p \varphi_n)^2 \nonumber \\
& & - s_n (\p\varphi_n)^2  -3 r_n L^{2n} s_n^2 - L^{2n} s_n^3
\end{eqnarray}
where $r_n(u) \equiv r(L^n (x-2L^n t)) =
 r (L^n u)$ and $q_n(u) \equiv
\frac{r'(L^n u)}{r(L^n u)}$.

The RG map (5) is now  defined by constructing a solution to (11,12)
on the time interval $[1,L^2]$ with
initial data $\varphi_n (u,1) = \varphi(u),
 s_n (u,1) = s(u)$ in a suitable Banach space,
\be
{\cal R}_n (\varphi,s) (\cdot) =
(\varphi_n (L\cdot, L^2), s_n (L\cdot, L^2)).
\en
One then wants to show that ${\cal R}_n$ maps the space
into another one so as to be able to iterate (5).

${\cal R}_n$ is studied by rewriting (11,12) as integral equations :
\begin{eqnarray}
\varphi_n (u,t) &=& \int dv R^{t-1}_n (u,v) \varphi_n
(v,1)
+ \int^{t-1}_0 d\tau \int dv R^\tau_n (u,v) M(v,t-\tau)
\non\\
&\equiv& (R^{t-1}_n\varphi_n(\cdot,1))(u)+{\cal M}_n(\varphi_n,s_n)(u,t)
\\
s_n(u,t) &=& \int dv Q_n^{t-1} (u,v) s_n (v,1)
+ \int^{t-1}_0 d\tau  \int dv Q^\tau_n (u,v) N(v,t-\tau)
\non\\
&\equiv& (Q^{t-1}_n s_n(\cdot,1))(u)+{\cal N}_n(\varphi_n,s_n)(u,t)
\end{eqnarray}
where $R^\tau_n , Q^\tau_n$ solve the linearised version of (11,12):
\BE
\dot{\varphi}_n & = &\p^2\varphi_n +
2 L^n(1+q_n) \p \varphi_n  \\
\dot{s}_n & = & \p^2 s_n + 2 L^n \p s_n
+ L^{2n} s_n (1-3 r^2_n)
\end{eqnarray}
Explicitely, see (2.12), (2.13), (2.19),
\begin{eqnarray}
R^\tau_n (u,v) &=& \exp(f_n (u) -f_n (v)) (\exp - \tau H_n)
(u,v) \\
Q^\tau_n (u,v) &=& \exp (- L^n u + L^n v) (\exp - \tau
\tilde{H}_n) (u,v)
\end{eqnarray}
where,
\be
f_n (u) \equiv f(L^n u), f(u) = -u-\log r(u)
\en
and
\BE
u &=& x-2L^n t,\\
v &=& y-2L^n,\\
t &=& \tau +1,
\EN
\begin{eqnarray}
H_n &=&- \p^2+ L^{2n} r^2_0 (L^n \cdot), \\
\tilde{H}_n &=& -\p^2+ 3 L^{2n} r^2_0 (L^n
\cdot).
\end{eqnarray}
$M, N$ collect the nonlinear terms in
(11,12) :
\begin{eqnarray}
M &=& 2L^n\p \varphi_n \left(
\frac{q_n+L^{-n}\p s_n/r_n}{1+s_n/r_n} - q_n \right), \\
 N &=& - r_n (\p \varphi_n)^2
- s_n (\p \varphi_n)^2 - 3 r_n L^{2n} s^2_n -
L^{2n} s^3_n.
\end{eqnarray}

Since the nonlinear terms involve $s_n, \p s_n, \varphi_n,
\p \varphi_n$, we shall solve equations (14,15) in a Banach
space ${\cal B}_n$ defined by a weighted $L^\infty$ norm on
$\varphi, s$ and their first derivative, with
$n$-dependent weights :
\be
\|(\varphi,s)\|^{(n)} = \sup_{t \in [1,L^2]} \;\; \|
\varphi (\cdot,t) \|^{(n)}_{1,t} +  \sup_{t \in [1,L^2]}
\;\; \| s(\cdot,t) \|^{(n)}_{2,t}
\en
with
\begin{eqnarray}
\|g\|^{(n)}_{1,t} &=& \| \frac{g}{h_{n,t}}\|_\infty +
\| \frac{g'}{{\tilde h}_{n,t}} \|_\infty \\
\|g\|^{(n)}_{2,t} &=& \| \frac{g}{w_{n,t}}\|_\infty +
 \| \frac{g'}{{\tilde w}_{n,t}}
\|_\infty
\end{eqnarray}
and where the weights are chosen to reflect the decay
rates of $\varphi,s$ :
\be
h_{n,t} (u) =
\left \{
\begin{array}{ll}
(1+\mid x \mid)^{-1-\delta} & u<0
\\
(1+u)^{-2-\delta} (1+L^n u)^{-1}(1+2L^nt)^{-1-\delta}
 & u\geq 0,
\end{array}
\right
\en
\be
{\tilde h}_{n,t}=L^{n(1+\delta)}{l}_{n}{ h}_{n,t}
\en
with
\be
l_n (u) = L^{-n (\frac{1}{4}-\delta)} + (1+ L^n |u|)^{-1/4}
\en
and
\be
w_{n,t} (u) = \left \{
\begin{array}{ll}
(L^n+\mid x \mid)^{-1-\delta}&u < 0
\\
(1+u)^{-2-\delta} e^{-L^nu}(L^n +
2 L^n t)^{-1-\delta}&u \geq 0,
\end{array}
\right
\en
\be
{\tilde w}_{n,t}=L^{n(1+\delta)}w_{n,t}
\en
where $u$ and $x$ are related by (21), i.e. $x$ is a function of $u$
and $t$ which makes the weights $t$-dependent.
For a function $g$
independent of $t$, we define $\|g\|^{(n)}_{1}$
as $\|g\|^{(n)}_{1,t}$ above
with $t =1$, and similarly for $\|g\|^{(n)}_{2}$.
 We write $h_n, w_n$
for $h_{n,1}, w_{n,1}$.

\vs{2mm}

\no {\bf Remark}.
These complicated weights can be understood as follows:
as we saw in (2.17), $\varphi$ contains two diffusive
waves, one in the
fixed frame, one carried by the front; the first term in
$h_{n,t}$ correspond to
the wave in the fixed frame, which we assume only
to decay (in space) like an integrable
power. The second term corresponds to the wave
carried by the front, which will
decay faster (in time), because the potential in
(24) creates a barrier at the origin (of
the front frame), so it decays essentially like
the solution of the heat
equation in $[0,\infty]$. This explains the
extra factor $(1+2L^nt)^{-1-\delta}$
(the role of $\delta$ will be discussed later).
However, to get this factor, we
need a slightly faster decay (in space) than in the
first term, so we put
$(1+u)^{-2-\delta}$. The factor $(1+L^n u)^{-1}$ is
 only due to the change of
variable (2.12) leading to the potential equation (recall (2.5).
 Finally, for later purposes,
we choose the coefficients so that $h_{n,t}(u)$ is
continuous.
Turning to (34),
we expect from (2.20) to have only a wave near the
 front. This is reflected in
the second term of (34), where again the factor
$e^{-L^nu}$ is related
to the change of variables leading to (2.20).
For the first term, the linear
analysis (2.20) would suggest a much faster decay.
 However, in (2.9), $s$ is
coupled to $\varphi$ by the nonlinear terms
and the first part of (34) is
 produced by this coupling. We will see that,
when we take
derivatives, $\varphi, s$ are multiplied by
 $L^n$, at least for $u$ (and $\tau$)
small. This is reflected by the factor
multiplying $h_{n,t}$ in (32), and by the
less refined factor in (35).

\vs{2mm}

The Theorem will be deduced from the

\vs{2mm}

\no{\bf Proposition 1}.{\it  There exists an $n_0>0$
such that for all $n\geq n_0$
\be
\|{\cal R}_n(\varphi,s)\|^{(n+1)}\leq L^{3\delta-1}\|(\varphi,s)\|^{(n)}
\en
if} $\|(\varphi,s)\|^{(n)}\leq L^{(3\delta-1)n}$.

\vs{2mm}

\no{\bf Remark}. The $n_0$ depends on the $\delta$ in (31-35)$ ($\delta$
is small, in particular $<{1\over 3}$). It is convenient to express
the smallness of the initial data in terms of $n_0$, that is, we start
the RG-iteration not at $n=0$, but at $n=n_0$. This is no loss of
generality, since we may scale the equations (2.7,8) by $\varphi(x,t)=
\phi(L^{-n_0}x,L^{-2n_0}t)$ and $s$ similarily. Then the norm (2.9) will
be bounded by the norm (28). The proposition then implies the Theorem
(with more detailed estimates on the $x$-dependence), for the discrete
times $L^{2n}$. For the remaining times the bounds below will give the
result.

\vs{2mm}

For the Proposition, we need to control the linear and the nonlinear
parts of ${\cal R}_n$. For the linear part, we have

\vs{2mm}

\no{\bf Lemma 1}. {\it Let $t\in[1,L^2]$. Then
\be
\| R_n^{t-1} \varphi \|^{(n)}_{1,t} \leq L^{2n\delta}
 \| \varphi \|^{(n)}_{1}\; ,\;\;
\| Q_n^{t-1} s \|^{(n)}_{2,t} \leq L^{2n\delta}
 \| s \|^{(n)}_{2}.
\en
Moreover,}
\BE
\| (R_n^{L^2-1} \varphi) (L \cdot) \|^{(n+1)}_{1} \leq
L^{-(1-2\delta)}  \| \varphi \|^{(n)}_{1}
\\
\| (Q_n^{L^2-1} s) (L \cdot) \|^{(n+1)}_{2} \leq
L^{-(1-2\delta)}  \| s \|^{(n)}_{2} ,
\EN

\vs{2mm}

\no{\bf Remark}. Note how the scaling in the RG produces
the contraction lacking in the Picard iteration. The $L^{2n\delta}$
in (37) is a brute force estimate: actually it could be replaced
by $Cn$. This divergence arises for very short times.

\vs{2mm}

For the nonlinear terms we need to use the contraction mapping
principle. We have

\vs{2mm}

\no{\bf Lemma 2}. {\it Let
\be
\|(\varphi,s) \|^{(n)} \leq  2L^{-n(1-5\delta)}.
\en
Then
\be
\|({\cal M}_n(\varphi,s),{\cal N}_n(\varphi,s))\|^{(n)} \leq
 L^{-n\delta}\|(\varphi,s) \|^{(n)}
\en
and the map ${\cal K}_n=({\cal M}_n,{\cal N}_n):{\cal B}_n
\rightarrow {\cal B}_{n}$ is a contraction in the ball {\rm (40)}:
\BE
\|{\cal K}_n(\varphi,s)-{\cal K}_n(\varphi',s')\|^{(n)}\leq
\theta\|(\varphi,s)-(\varphi',s')\|^{(n)}\non
\EN
for $\theta<1$.  }

\vs{2mm}

Thus, by (37) and Lemma 2, if we call $(\varphi^0_n,s^0_n)$
the linear terms in (14,15), we can solve
these equations by the contraction mapping principle applied to
a ball of radius $L^{-n(1-5\delta)}$ around $(\varphi^0_n,s^0_n)$.
Indeed, using (36) inductively and (37), one gets
$\|(\varphi^0_n,s^0_n) \|^{(n)} \leq  L^{-n(1-5\delta)}$.
Then (41), (38) and (39)
yield (36) i.e. the Proposition, and (36), (37)
and (41) the Theorem (by redefining $\delta$).
Hence, we need only to prove Lemmas
1 and 2.

\vs{2mm}

In the course of the proofs,
we shall write  $C$ or
$c$ to denote suitable constants which may vary from place
to place (even in the same formula)
but do not depend on $L$ or $n$ while $C(L)$
depends on $L$ but not on $n$.
$L$ will be fixed, but chosen large enough so that we may
repeatedly control constants by writing, e.g.,
 \be
C \leq L^\delta
\en
$L$-dependent constants are controlled by
\be
C(L) \leq L^{\delta n}
\en
since $n\geq n_0$.
This explain the proliferation, in the proofs,
of powers of $L^{\delta n}$. They are also used to control factors
like $C(L) n$
.The logic in the choice of $L$ and $n_0$ is as
follows : given $\delta > 0$, small we
choose $L$ large enough so that (42) holds for
all the $L$-independent constants entering the proofs
and then we choose $n_0$ so that (43) holds when
 $n \geq n_0$, for all $L$-dependent constants.

In the proofs, we shall need some properties
of the ``front" $r$, which
follow from an analysis of (2.4). One knows
that $r (u) \simeq ue^{-u}$ for
$u \rightarrow +\infty$ while $1- r (u) \simeq e^{-cu}$ for
$u \rightarrow -\infty$. More precisely,
we have for the function $f(u)$
defined in (20),
\be
\mid f(u) +u\mid \leq \lambda e^{cu}
\en
\be
\mid f'(u) +1\mid \leq e^{cu}
\en
for $u\leq 0$. Moreover, we can choose the
origin of the coordinates
so that $\lambda$ in (44) is as small as we wish.
This possibility will be used
later. Besides, we have
\be
\mid f(u) +\log(1+u)\mid \leq c
\en
\be
\mid f'(u) \mid \leq \frac{c}{1+u}
\en
for $u\geq 0$. Finally, we have (see e.g. (45, 47))
\be
\mid q(u)\mid =\mid\frac{r'(u)}{r(u)} \mid \leq {c}
\en.

\vs{5mm}

\no{\bf 3.2. Proof of Lemma 1.}

\vs{5mm}

We start by proving (38). For this it is enough to show:
\be
 \mid \int dv R_n^{L^2-1}
(Lu',v) h_n (v) \mid \leq  \ha L^{-(1-2\delta)} h_{n+1} (u')
\en
and
\BE
\mid \frac{d}{d u'} \int dv R_n^{L^2-1}
(Lu',v) \varphi (v) \mid
\leq  \ha L^{-(1 - 2\delta)}
L^{{(n+1) }(1+\delta)}l_{n+1} (u') h_{n+1} (u')
(\|\frac{\varphi}{{ h}_n}\|_\infty
+\|\frac{\varphi'}{{\tilde h}_n}\|_\infty)
\EN
We use primes to denote variables on the scale $n+1$
and unprimed ones for scale $n$. Note that $u'=x' -2L^{n+1}t'$,
so we have
$x=Lx'$ for $u=Lu', t=L^2 $, i.e. $t'=1$ .

To prove (49, 50), we need some
properties of $R^{\tau} _n$ : we shall use the path space
representation.
\be
(\exp - {\tau}  H_n) (u,v) = \int d \mu^{\tau} _{u,v} (\omega) e^{-
\int^{\tau} _0 V_n(\omega(s)) ds}
\en
where $d \mu^{\tau} _{u,v}$ is the Brownian bridge going in
time ${\tau} $ from $u$ to $v$, so that
$\int d \mu^{\tau} _{u,v} (\omega)
= \frac{e^{-(u-v)^2 /4{\tau} }}{\sqrt {4 \pi {\tau} }}$
 and $V_n(\cdot) =
L^{2n} r^2 (L^n \cdot).$
Thus the potential has a high and sharp barrier around
zero which is repulsive on the negative real axis. To get
simple estimates we introduce
\be
\epsilon_n = k n L^{-n}\log L
\en
where $k$ is chosen large so that $V_n$ is essentially
equal to $0$ or to $L^{2n}$ outside the interval $I_n = [-
\epsilon_n, + \epsilon_n]$. In Lemma 3 and later,
we have bounds where
the quantity $L^{-np}$ enters. Here, $p={\cal O}(k)$ can
 be taken large by choosing
$k$ in (52) large, and the quantities bounded by
 $L^{-np}$  will
turn out to be negligeable.
The bounds based on the path space representation
(51) will be stated in a
series of Lemmas, whose proofs will be given in the Appendix.

\proclaim Lemma 3.
For $0\leq {\tau}  \leq L^2-1$ and $n\geq n_0$, we have
\begin{enumerate}
\item[a)]
Let $u \geq 0$. If $v \geq - \epsilon_n$, then
\be
\exp(-\tau H_n)(u,v) \leq D^{\tau} _{-2 \epsilon_n} (u,v) +
{L^{-np}} e^{- \frac{c}{\sqrt{\tau}} (u+v)}
\en
where $D^{\tau}_a (u,v)$ is the Dirichlet kernel with barrier
at a :
\be
D^{\tau}_a (u,v) = H^{\tau} (u-v) - H^{\tau} (u+v-2a)
\en
for $u,v > a$, with
\be
H^{\tau} (u) = \frac{e^{-\frac{u^2}{4 \tau}}}{\sqrt{4 \pi {\tau}}}
\en
being the heat kernel.
Moreover,
\be
\int^{- \epsilon_n}_{- \infty} dv
R^{\tau}_n (u,v)(1+\frac{|v|}{\sqrt{\tau}})
 \leq L^{-np}
e^{-c \frac{u}{\sqrt{\tau}}}(1 + L^n u)^{-1}
\en
\item[b)]
Let $ u \leq 0$. If $v \leq - \epsilon_n$, then
\be
R^{\tau}_n (u,v) \leq \frac{C}{\sqrt{{\tau}}}\exp
(-\frac{(u-v+2L^n{\tau})^2}{4{\tau}})+ L^{-np} e^{-
\frac{c}{\sqrt{\tau}}(\mid v \mid + \mid x \mid \chi(x<0))}
\en
and for all $t\in [1,L^2]$,
\BE
&\int^\infty_{- \epsilon_n}  dv R^{\tau}_n (u,v)
h_{n, t- \tau}(v) (1+L^{-n}\frac{|u-v|}{\sqrt{\tau}}) \leq \non \\
&h_{n, t}(u) (L^{-\frac{n}{3}}+
L^{\frac{n\delta}{4}} l_n(u) \chi(\tau \leq
L^{-\frac{n}{2}})) e^{-
\frac{c}{\sqrt{\tau}} \mid x \mid \chi(x<0)}
\EN
\end{enumerate}
\par

\no {\bf Remark}. Except for (58), these bounds are intuitively obvious:
in (53) correction to the Dirichlet kernel comes from excursions to the
high potential region, whence the $L^{-np}$. In (56) such excursions
have to take place and in (57) the leading term comes from the heat
kernel, since by (21-23),
\be
u-v+2L^n{\tau}=x-y.
\en

\vs{2mm}

Using Lemma 3 for $u=Lu'$ and $\tau =L^2 -1$,
we can prove (49).
Consider first $u' \geq 0$, and write
the integral in (49) as:
\be
 \int dv = \int^\infty_{-\epsilon_n} dv +
\int^{-\epsilon_n}_{- \infty} dv
\en
The second integral is trivial by (56), so consider the
first one. Use (18, 53), and the bounds
\be
e^{-f_n(v)} h_n(v)\leq CL^{-n(1+\delta)}(1+v)^{-2 - \delta}
\en
for $v\geq -\epsilon_n$, and
\be
e^{f_n(Lu')} \leq C(1+L^{n+1} u')^{-1}
\en
for $u'\geq 0$
,which follow from (31, 46, 44).
 We see then that all we have to show is
\be
\int^\infty_{-\epsilon_n} dv D^{L^2-1}_{-2\epsilon_n}
(Lu',v) (1+v)^{-2 - \delta} \leq
 L^{-2+ \frac{\delta}{2}} (1+u')^{-2-\delta}
\en
Indeed, the second term
in (53) is trivial and, in (63), we
write $ L^{-2+\frac{\delta}{2}} =
L^{-\frac{\delta}{2}}L^{-(1-2\delta)} L^{-(1+\delta)}$;
$ L^{-(1-2\delta)}$ occurs in (49),
 $L^{-(1+\delta)}$ is
used, with $L^{-n(1+\delta)}$ in (61), to obtain the factor
$(1+2L^{n+1})^{-1-\delta}$ in $h_{n+1}(u')$ (for $u'\geq 0$)
and $L^{-\frac{\delta}{2}}$ is used to control
constants, as in (42).

(63) follows from (54), the definition of $D$:
for $u'\geq k\log L$, with $k$ large, we
use simply
\be
D^{L^2-1}_{-2 \epsilon_n} (Lu',v) \leq
 H^{L^2-1}(Lu'-v)
\leq \frac{C}{L} e^{-c
\mid u'-\frac{v}{L} \mid}
\en
and
\be
L^{-1}\int^\infty_{-\epsilon_n} dv e^{-c
\mid u'-\frac{v}{L} \mid}
 (1+v)^{-2 - \delta} \leq C  (1+Lu')^{-2-\delta}
\en
with $(1+Lu')^{-2-\delta} \leq  L^{-2} (1+u')^{-2-\delta}$
for such $u'$.

For $u' \leq
 k\log L$, we use
\be
D^\tau_a (u,v) = \mid \int^{u+v-2a}_{u-v} dz
\frac{d}{dz} H^\tau (z)\mid \leq c\tau^{-1}
(\mid v \mid + \mid a \mid)
\en
to get
\be
D^{L^2-1}_{-2 \epsilon_n} (Lu',v)
\leq \frac{c}{L^2}(|v|+
\epsilon_n)
\en
which gives (63).

For $u'\leq 0$, we use (60) in (49),
insert (57) in the second integral and observe that, for
$\tau =L^2-1$, $u=Lu'$, (59) becomes:
$ Lu' - v +2L^n(L^2 -1)= Lx'-y $. Using the second inequality
in (64) on $H^{L^2-1}(Lx'-y)$, the first term of the
RHS of (57) gives a bound on
$\int_{-\infty}^{-\epsilon_n} dv R^{L^2-1} (u,v) h_n(v) $
of the form
\be
\frac{C}{L} \int_{v \leq - \epsilon_n} dy e^{-c
\mid x'-\frac{y}{L} \mid} (1+ \mid y \mid)^{-1-\delta}
\leq CL^{-1}
(1+ \mid x' \mid)^{-1-\delta}
\en
which yields a contribution to the RHS of (49)
for that part of the integral.
The second term in (57) gives another contribution to (49):
we get exponential decay for $x'\leq 0$ and, for $x'\geq 0$,
 we may use part of
$L^{-np}$ to get the factor $(1+x')^{-1-\delta}$
in $h_{n+1}(u')$, since $x'\leq 2L^{n+1}$ for $u'\leq 0$.

Finally,
for $\int_{v>- \epsilon_n} dv$, we use (58),
where only the term with
$L^{-\frac{n}{3}}$ in the RHS contributes,
since $t=L^2$, $\tau= L^2 -1$. Moreover,
for $u=Lu'$ (of any sign),
\be
h_{n, L^2 }(u) \leq  h_{n+1}(u')
\en
 We use, as in (43), the factor
$L^{-\frac{n}{3}}$ in (58)
to bound constants,
 and to give
$L^{-(1-2\delta)}$ in (49).

Now, we prove (50). Let us consider first $u'\geq 0$.
We use the following formula, valid for any continuous,
piecewise differentiable function $g$ with
$\|g\|_\infty + \|g'\|_\infty < \infty$,
\BE
& &\frac{d}{du} (R^{\tau}_n g) (u) = \frac{d}{du}
\int dv R^{\tau}_n (u,v)
g (v) \non \\
 &=& L^n f' (L^n u) \int dv R^{\tau}_n (u,v) g (v) \non \\
 &-& \int dv e^{f_n (u)-f_n(v)} \int d \mu^{\tau}_{u,v}
 (\omega) (e^{-
\int^{\tau}_0 V_n (\omega(s))ds} L^{3n}
\int^{\tau}_0 V' (L^n \omega (s))ds)g(v)
\non \\
 &-& L^n \int dv R^{\tau}_n (u,v) f' (L^n v) g(v) \non \\
&+&\int dv R^{\tau}_n
(u,v) g' (v) dv
\EN

To prove (70), use the definition of $R^{\tau}_n$ (18),
the path space formula
(51) and the identity, coming from the translation
invariance of $d \mu^{\tau}_{u,v}$ :
\be
\frac{d}{da} \int d \mu^{\tau}_{u+a,v+a}(\omega)
 e^{-\int^{\tau}_0 V_n
(\omega(s)-a)ds} =0
\en
which implies
\BE
(\frac{d}{du}+\frac{d}{dv}) (e^{-{\tau}H_n}) (u,v) =
 - \int d \mu^{\tau}_{u,v} (\omega) (e^{-
\int^{\tau}_0 V_n (\omega(s))ds} L^{3n}
\int^{\tau}_0 V' (L^n \omega (s))ds)
\EN
Integration by parts of the $d/dv$ gives the
last two terms of (70).

Consider each term of (70), with $g=\varphi$,
 ${\tau}=L^2-1$ and $u=Lu'$ $(\geq 0) $ so that
$\frac{d}{du'}=L\frac{d}{du}$ .
We shall use
 the fact that in  (50), we have a factor
$L^{{(n+1) }(1+\delta)} l_{n+1} (u') $
while all the terms of (70), except the last
one, involve an integration with $\varphi (v)$
instead of its derivative.

For the first term in (70), we use (47), which yields
\be
L^{n+1} |f' (L^{n+1}u')| \leq \frac{L^{n+1}}{1+L^{n+1} u'}
\leq L^{{n+1 }} l_{n+1} (u').
\en
This, combined with (49), gives a contribution to the RHS of (50).

For the last two terms of  (70), it is enough to prove that:
\be
\int dv R_n^{L^2-1} (Lu',v) (1+L^n |v|)^{-{1\over 4}}
 h_n(v) \leq
C L^{-{n \over 4} } h_{n+1}(u').
\en
for $u'\geq 0$. Indeed, as in (73),  $|f' (L^{n}v)| \leq
(1+L^n |v|)^{-{1\over 4}}$ and
we
can bound $\varphi'$ in terms of $l_n h_n$.
We use (49) for the first term in $l_{n} $. Then, we write
 $ L^{-{n\over 4} }\leq L^{-{n} \delta}l_{n+1} $.
The factor $C$ in (74) and $L^{-(1-2\delta)}$
in (50) can be controlled using
$L^{-{n} \delta}$ here and (43). To prove (74),
we follow the proof of
(49) for $u'\geq 0$ and use, instead of (63),
\be
\int^\infty_{-\epsilon_n} dv
D^{L^2-1}_{-2\epsilon_n} (Lu',v) (1+v)^{-2 - \delta}
(1+L^n |v|)^{-{1\over 4}}
\leq C L^{-{n \over 4} }(1+u')^{-2-\delta}
\en
which is easy to prove, using (63) for $v\geq 1$
and bounding $D^{L^2-1}_{-2\epsilon_n}
(Lu',v) (1+v)^{-2 - \delta}$ by $Ce^{-cu'}$ for $v\leq 1$.

For the second term in (70), it is enough to show:
\BE
&L\int dv e^{f_n (Lu')-f_n(v)} k_n (Lu',v,L^2 -1) h_n(v)
\leq C(L) L^{{2n \over 3}} h_{n+1}(u') \non \\
& \leq L^{-n\delta }L^{(n+1)(1+\delta) }
{ l}_{n+1}(u'){ h}_{n+1}(u')
\EN
where
the second inequality follows from (33, 43),
and we used the notation
\be k_n (u,v,{\tau}) =
\int d \mu^{{\tau}}_{u,v}(\omega)
\left( e^{-\int^{\tau}_0 V_n (\omega(s))ds} L^{3n}
\int^{\tau}_0 |V' (L^n \omega (s))| ds \right)
\en
The first inequality of (76) follows from parts
a and b of the following Lemma, for $\tau =L^2 -1$,
and (61, 62).

\proclaim Lemma 4.
With the notation (77), we have, for
$0\leq {\tau} \leq L^2-1$
 and $n\geq n_0$:
\begin{enumerate}
\item[a)]
for $u \geq 0$,
\be
\int^\infty_{-\epsilon_n} dv k_n (u,v,{\tau})
\leq  e^{- \frac{c}{L} u}
(L^{\frac{2n}{3}}+ L^{n(1+\frac{\delta}{2})}
 l_n(u) \chi(\tau \leq
L^{-\frac{n}{2}}))
\en
\item[b)]
for $u \geq 0$,
\be
\int^{-\epsilon_n}_{-\infty} dv e^{-f_n(v)}k_n (u,v,{\tau})
\leq L^{-np} e^{- \frac{c}{\sqrt{\tau}} u}
(1 + L^n u)^{-1}
\en
\item[c)]
for $u\leq0$,
\BE
&\int^\infty_{-\epsilon_n } dv \exp (f_n (u)-f_n(v))
k_n (u,v,{\tau}) h_{n, t-\tau}(v) \non \\
&\leq
h_{n,t}(u) (L^{n(\frac{2}{3}+ \frac{\delta}{4})}+
 L^{n(1+\frac{\delta}{2})} l_n(u) \chi(\tau \leq
L^{-\frac{n}{2}}))
\EN
\item[d)]
for $u\leq 0$,
\be
\int^{-\epsilon_n }_{-\infty} dv\exp(f_n (u)-f_n(v))
k_n (u,v,{\tau}) \leq
L^{-np} e^{- \frac{c}{\sqrt{\tau}} \mid x \mid \chi (x<0)}
\en
\end{enumerate}
\par

Now we prove (50) for $u'\leq0$. We use a different
formula for $\frac{d}{du} (R^{\tau}_n g) (u)$. Write
\be
R^{\tau}_n(u,v) =\frac{1}{\sqrt {4 \pi {\tau}}}
e^{f_n(u)+L^n u-(f_n(v)+L^n v) - (u-v+2L^n {\tau})^2/4{\tau}}
< e^{-\int^{\tau}_0 V (\omega(s)) ds} >^{\tau}_{u,v}
 e^{L^{2n} {\tau}}
\en
where we use the normalised expectation value
\be
<G>^{\tau}_{u,v} = (\int d \mu^{\tau}_{u,v}
 (\omega))^{-1} \int d \mu^{\tau}_{u,v}
(\omega) G(\omega)
\en
Taking derivatives, we have, for $g$ as in (70),
\BE
& &\frac{d}{du} \int dv R^{\tau}_n (u,v) g(v) =
L^n (f' (L^nu)+1) \int dv R^{\tau}_n (u,v) g (v) \non \\
&-&\int dv e^{f_n (u)-f_n(v)} \int
d \mu^{\tau}_{u,v} (\omega) (e^{-
\int^{\tau}_0 V_n (\omega(s))ds} L^{3n}
\int V' (L^n \omega (s))ds)g(v)  \non \\
&-& L^n \int dv (f'(L^n v)+1) R^{\tau}_n (u,v) g (v)
+\int dv R^{\tau}_n (u,v) g' (v)
\EN
where the second term comes from the
application
of (72) (to the normalised expectation value,
which obviously satisfies (71)) and the
last two terms are produced by the integration
 by parts of $\frac{d}{dv}$. The term
coming from  \be \frac{d}{du}
 e^{-(u-v+2L^n {\tau})^2/4{\tau}} \en
is cancelled by a similar term
produced by the integration by parts.

Now we bound each of these terms for $g=\varphi$,
$u=Lu'\leq 0 $ and ${\tau}=L^2-1$.
For the first term, it is enough to use (49)
and (45):
\be
L^{n+1} \mid f' (L^{n+1} u') +1 \mid \leq
L^{n+1}l_{n+1}(u')
\en
For the last two terms, as in (70),
it is enough to use (49) and:
\be
\int dv R_n^{L^2-1} (Lu',v) (1+L^n |v|)^{-\frac{1}{4}}
 h_n(v)  \leq
C L^{-{n \over 4} } h_{n+1}(u')
\en
for $u'\leq 0$. For this, we follow the proof of (49) for $u' \leq 0$,
with the only change that, instead of (68), we use
\be
\int_{v \leq - \epsilon_n} dy e^{-c
\mid x'-\frac{y}{L} \mid} (1+ \mid y \mid)^{-1-\delta}
(1+L^n |v|)^{-\frac{1}{4}}
\leq CL^{-\frac{n}{4}}
(1+ \mid x' \mid)^{-1-\delta}
\en
This is easy to show: for $|v| \geq 1$, use (68) and, for
$|v| \leq 1$, use the bound
$$e^{-c
\mid x'-\frac{y}{L} \mid} (1+ \mid y \mid)^{-1-\delta}
\leq C (1+ \mid x' \mid)^{-1-\delta}.$$
Note that in (58) we have a factor $L^{-\frac{n}{3}}$
(for $\tau = L^2 -1$, $t=L^2$),
so (87) holds
trivially, for that part of the integral,
using (69).
Finally,
for the second term in (84), we prove (76), using
parts c and d of Lemma 4: in (80), we have only the
first term in the RHS, because
$\tau=L^2-1$ ($t=L^2$), and we use
(69).  In (81),
we use, as before, part of $L^{-np}$ to get the factor
$(1+|x'|)^{-1-\delta}$
in $h_{n+1}(u')$ (since $x'\leq 2L^{n+1}$ for $u'\leq 0$).
This completes the proof of (50) and, therefore, of (38).
To prove the first estimate in (37), it is enough to show:
\be
 \mid \int dv R_n^{\tau}
(u,v) h_n (v) \mid \leq  L^{n\delta} h_{n, \tau +1} (u)
\en
and
\BE
\mid \frac{d}{d u} \int dv R_n^{\tau}
(u,v) \varphi (v) \mid   \leq
L^{{n }(1+2\delta)}l_{n} (u) h_{n, \tau +1} (u)
(\|\frac{\varphi}{{ h}_n}\|_\infty
+\|\frac{\varphi'}{{\tilde h}_n}\|_\infty)
\EN

We follow the proof of (49, 50) (Lemmas 3 and 4
 hold for all $\tau$).
To prove (89), we use
\be
\int^\infty_{-\epsilon_n} dv D^{\tau}_{-2\epsilon_n}
(u,v) (1+v)^{-2 - \delta} \leq  C(L) (1+u)^{-2-\delta}
\en
instead of (63), and, instead of (68),
\be
 \frac{1}{\sqrt\tau}\int_{v \leq - \epsilon_n} dy e^{-c
\frac{|x_\tau-y|}{\sqrt\tau}} (1+ \mid y \mid)^{-1-\delta}
\leq C(L)
(1+ \mid x_\tau \mid)^{-1-\delta}
\en
where, see (21-23),
\be
x_\tau= u+ 2L^n(\tau +1)
\en
is the variable entering $h_{n, \tau +1}$.
When we use (58) (with $t=\tau +1$),
we have to consider the
 second term in the RHS, for
small $\tau$. However, previously, we used
 $L^{-\frac{n}{3}}$ only to control
$C(L)$ (see the remark following eq. (69)).
 So, here, we use $l_n \leq 2$,
$C(L) L^{\frac{n\delta}{4}}\leq L^{\frac{n\delta}{2}}$
which is enough, given the factor
$L^{n\delta}$ in the RHS of (89).

To prove (90), we use
\be
\int dv R^{\tau}_n (u,v) (1+L^n |v|)^{-{1\over 4}}
 h_n(v) \leq
L^{n\delta}  l_n(u)  h_{n, \tau +1}(u)
\en
for all $u$,
instead of (74) and (87), which
is proven like (74, 87), using
$$
\int^\infty_{-\epsilon_n} dv
D^{\tau}_{-2\epsilon_n} (u,v) (1+v)^{-2 - \delta}
(1+L^n |v|)^{-{1\over 4}}
\leq C(L) L^{-{n \over 4} }(1+u)^{-2-\delta}
$$
for $u\geq 0$, instead of (75),
$$
 \frac{1}{\sqrt\tau}\int_{v \leq - \epsilon_n} dy e^{-c
\frac{|x_\tau-y|}{\sqrt\tau}} (1+ \mid y \mid)^{-1-\delta}
(1+L^n |v|)^{-\frac{1}{4}}
\leq C(L) L^{-\frac{n}{4}}
(1+ \mid x_\tau \mid)^{-1-\delta}
$$
for $u\leq 0$, instead of (88), and
$L^{-\frac{n}{4}} \leq l_n(u) $. Finally, we use
\be
\int dv e^{f_n (u)-f_n(v)} k_n (u,v,\tau) h_n(v)
\leq C(L)
L^{n(1+\frac{\delta}{2})} l_n(u)  h_{n, \tau +1}(u)
\en
for all $u$, instead of (76);
 (95) follows from Lemma 4 (for $t=\tau +1$ in (80)) and (61, 62).

Finally we shall prove (39) and the second
estimate in  (37), and, as before, concentrate on (39).
We shall take advantage of the factor 3 in (25),
as opposed to 1 in (24).
This factor means that the potential barrier,
for $u\leq 0$, is higher, and,
as a consequence,
we shall no longer have to deal with the front
 in the laboratory frame
(compare the first terms in (2.17) and (2.20)).
To prove (39), it is enough to
show (remember that ${\tilde w}$ is a multiple
of $w$, see (35)):
\be
 \mid \int dv Q_n^{L^2-1}
(Lu',v) w_n (v) \mid \leq  \ha L^{-(1-2\delta)} w_{n+1} (u')
\en
and,
\be
\mid \frac{d}{d u'} \int dv Q_n^{L^2-1}
(Lu',v) s (v) \mid
\leq  \ha L^{-(1 - 2\delta)}
w_{n+1} (u')
(L^{(n+1)(1+\delta)}\|\frac{s}{w_n}\|_\infty
+ L^{1+\delta}\|\frac{s'}{w_n}\|_\infty)
\en
Instead of Lemma 3, we use

\proclaim Lemma 5.
For $0<{\tau} \leq L^2 - 1$ and $n\geq n_0$, we have
\begin{enumerate}
\item[a)]
for $u \geq 0$, if $ v \geq - \epsilon_n$,
\be
e^{-\tau{\tilde H}_n} (u,v)
\leq   D^{\tau}_{-2\epsilon_n} (u,v) +
 {L^{-np}} e^{- \frac{c}{\sqrt{\tau}}
(u+v)}
\en
and
\be
\int^{-\epsilon_n}_{- \infty} dv Q^{\tau}_n (u,v)
(1+\frac{|v|}{\sqrt{\tau}})
\leq L^{-np}
e^{-\frac{c}{\sqrt{\tau}}u}
e^{-L^n u}
\en
\item[b)]
for $u\leq 0,$ if $ v \leq -\epsilon_n$,
\be
Q^{\tau}_n (u,v) \leq \frac{C}{\sqrt {\tau}}
\exp (-c{\tau}L^{2n} -\frac{(u-v+2L^n{\tau})^2}{4{\tau}})
+ L^{-np}
e^{-\frac{c}{\sqrt {\tau}} (\mid u \mid + \mid v \mid)}
\en
and
\be
\int_{-\epsilon_n}^\infty dv
Q^{\tau}_n (u,v)w_{n,t-\tau} (v)(1+\frac{|v|}{\sqrt{\tau}})
\leq w_{n,t} (u) (L^{-\frac{n}{3}} +
\chi (\tau \leq L^{-\frac{n}{2}}))
e^{-\frac{c}{\sqrt {\tau}} \mid u \mid}
\en
\end{enumerate}
Moreover, the same bounds hold, with different
constants, when $3$ in (25)
is replaced by any $\alpha >1$.
\par

Let us prove (96). For $u'\geq 0$, we can use part
 a of Lemma 5, exactly
as we used part a of Lemma 3 to prove (49),
the only difference
being that (61, 62) are replaced by
\be
e^{L^n v} w_n(v)\leq c L^{-n(1+\delta)}(1+v)^{-2 - \delta}
\en
for $v\geq -\epsilon_n$, which follows
immediately from (34), and
\be
e^{-L^{n+1}u'}
\en
for $u'\geq 0$.
The proof of (96) for $u'\leq 0$
is easy, given part b of Lemma 5: for $\tau
=L^2 -1$, even the first term of (100) is small.
We use $w_{n, L^2} (Lu') \leq L^{1+\delta}w_{n+1} (u')$
instead of (69).

To prove (97), we use the following formula,
valid for $g$ as in (70)
and any $u$:
\BE
\frac{d}{du} (Q^{\tau}_n g) (u) &=& \frac{d}{du}
\int dv Q^{\tau}_n (u,v)
g (v) \non \\
 &=& - \int dv e^{-L^n u + L^n v}
\int d \mu^{\tau}_{u,v} (\omega) (e^{-3
\int^{\tau}_0 V_n (\omega(s))ds} L^{3n} 3\int^{\tau}_0
V' (L^n \omega (s))ds)g(v)
\non \\
 &+&   \int dv Q^{\tau}_n
(u,v) g' (v) dv
\EN
This is proven like (70), by noticing that
$f(u)$ in $R$ is replaced
by $-u$ here so that the first and
the third terms in (70) cancel
each other. To bound the second term of (104),
just use (96), since
${\tilde w}$ is a multiple of $w$. For the first term,
observe that, by (48), $|V'|\leq cV$, so that
\be
3 L^{3n} \int^{\tau}_0|V' (L^n \omega (s))|ds
\leq C L^n e^{
\int^{\tau}_0 \ha V_n (\omega(s))ds}
\en
Then, we use the last statement in Lemma 5, with
$\alpha =\frac{5}{2}$, and use the factor
$L^{(n+1)(1+\delta)}$ in (97) to control $L^n$ in (105).
Again, the proof of (37) is a simple modification
of the proof of (39).

\vspace*{5mm}
{\bf 3.1. Proof of Lemma 2.}
\vspace*{5mm}

We need to show that
\be
\sup_{t\in [1,L^2]} \| \int^{t-1}_0 d{\tau} \int dv
R^{\tau}_n(u,v)M(v,t-{\tau}) \|^{(n)}_{1,t} \leq
 L^{-n\delta}\|(\varphi,s) \|^{(n)}
\en
and
\be
\sup_{t\in [1,L^2]} \| \int^{t-1}_0 d{\tau} \int dv
Q_n^\tau(u,v)N(v,t-{\tau}) \|^{(n)}_{2,t} \leq
L^{-n\delta}\|(\varphi,s) \|^{(n)}
\en
for $(\varphi,s)$ with
\be
\|(\varphi,s) \|^{(n)} \leq 3 L^{-n(1-5\delta)}
\en
($M$, $N$ are defined in (26, 27)).

The proof that the nonlinear terms define a
contraction is similair.

To prove (106), we shall prove, for all $t\in [1,L^2]$,
\be
 \mid \int^{t-1}_0 d{\tau} \int dv R_n^{\tau}
(u,v) M (v,t-{\tau}) \mid \leq
 L^{-n\delta} h_{n, t} (u)
\|(\varphi,s) \|^{(n)}
\en
and
\BE
\mid \frac{d}{d u} \int^{t-1}_0 d{\tau}\int dv R_n^{\tau}
(u,v) M (v,t-{\tau})  \mid   \leq
L^{-n\delta} L^{n(1+\delta)} l_{n} (u) h_{n, t} (u)
\|(\varphi,s) \|^{(n)}
\EN

Let us start with the proof of (109).
We use the bound
$r^{-1}_n \leq c(1+e^{L^nv})$ to get  from (34)
that $w_n r^{-1}_n \leq L^{-n}$, and, from (108),
$$
\frac{|s_n|}{r_n} \leq L^{-n(2-{\cal O}(\delta))}
$$
where we use throughout the proof ${\cal O}(\delta)$
 because we do not
need to keep track of the constants multiplying $\delta$.
We also have
$$
\mid \frac{1}{1+\frac{s_n}{r_n} } -1\mid \leq
L^{-n(2-{\cal O}(\delta))}
,$$
$$
\frac{|\p s_n|}{r_n} \leq L^{-n(1-{\cal O}(\delta))}$$
and
$$
\mid \p \varphi_n (v,t-{\tau}) \mid \leq L^{n(1+{\cal O}(\delta))}
l_n(v) h_{n, t-\tau}(v)\|(\varphi,s)
\|^{(n)}. $$
We also know from (48) that $|q_n| \leq c$, so , altogether,
\be
\mid M (v,t-{\tau})\mid
\leq L^{n{\cal O}(\delta)} l_n(v)
h_{n, t-\tau} (v)\|(\varphi,s) \|^{(n)}.
\en
Then, (109) follows by inserting (111) in (109),
using the bound
\be
\int^{t-1}_0 d{\tau}  \int dv R_n^\tau (u,v) l_n (v)
h_{n, t-\tau} (v) dv \leq L^{2n
\delta} L^{-\frac{n}{4}} h_{n, t} (u)
\en
and using $L^{-{n \over 4}} $ to control
$L^{n{\cal O}(\delta)}$ and to give
$L^{-n\delta}$ in (109); (112) is proven by combining
\be
\int^{t-1}_0 d{\tau}  \int dv R_n^\tau (u,v)
h_{n, t-\tau} (v) dv \leq
 L^{n\delta} h_{n, t} (u)
\en
which is proven like (89), and
\be
\int^{t-1}_0 d{\tau}  \int
dv R_n^\tau (u,v) (1+L^n|v|)^{-{1\over 4}}
h_{n, t-\tau} (v) dv \leq L^{n\delta}
 L^{-\frac{n}{4}} h_{n, t} (u)
\en
which is proven like (94),
 the only difference being that, when we use (58),
we have
$\int_0^{t-1}d\tau \chi(\tau \leq L^{-{n\over2}})\leq
 L^{-{n\over2}}$ for the second term of (58),
 so that the
result has a bound with $ L^{-\frac{n}{4}}$, as in (87).

To prove (110), we would like to use (70, 84).
However, we cannot
use integration by parts as in (70, 84) because we have
$\p\varphi, \p s$ in the nonlinear terms
and the norms (29, 30)
give no control over the second derivatives
 of $\varphi$ or $s$. We proceed as follows:
write, for $u\geq 0$,
\be
R^{\tau}_n(u,v) =\frac{1}{\sqrt {4 \pi {\tau}}}
e^{f_n(u) -f_n(v) - (u-v)^2/4{\tau}}
< e^{-\int^{\tau}_0 V (\omega(s)) ds} >^{\tau}_{u,v}
\en
Taking derivatives, we have
\BE
& &\mid \frac{d}{d u} \int^{t-1}_0 d{\tau}\int dv R_n^{\tau}
(u,v) M (v,t-{\tau})  \mid \leq \non \\
& &L^{n{\cal O}(\delta)} \|(\varphi,s) \|^{(n)}
\left( \mid L^n f' (L^nu)\mid \int^{t-1}_0 d{\tau}
\int dv R^{\tau}_n (u,v) l_n(v) h_{n, t-\tau} (v) \right. \non \\
&+& \int^{t-1}_0 \int dv e^{f_n (u)-f_n(v)} k_n(u,v,\tau)
l_n(v) h_{n, t-\tau} (v)  \non \\
&+& \int^{t-1}_0 d{\tau}
\int dv \frac{\mid u-v \mid}{2\tau}
R^{\tau}_n (u,v) l_n(v) h_{n, t-\tau} (v) \non \\
 &+& \int^{t-1}_0 d\tau \int dv
\frac{1}{\sqrt {4 \pi {\tau}}}
e^{f_n(u) -f_n(v) - (u-v)^2/4{\tau}}\non \\
& &\left. \mid \frac{d}{dv}< e^{-\int^{\tau}_0
 V (\omega(s)) ds} >^{\tau}_{u,v}\mid
 l_n(v) h_{n, t-\tau} (v) \right)
\EN
where we use (72) for the normalised expectation value
 and then take absolute values and
use (111) .
 Now, observe that
$\frac{d}{dv}< e^{-\int^{\tau}_0 V (\omega(s)) ds} >^{\tau}_{u,v}$
 is positive by the FKG inequality \cite{F,Si} :
$V$ is decreasing so $e^{-
\int V}$ is increasing and the Wiener measure
satisfies FKG inequalities, so
$< e^{-\int^{\tau}_0 V (\omega(s)) ds} >^{\tau}_{u,v}$
increases if the ''boundary conditions" $u$ or
$v$ increases. Therefore, we may take
out the absolute values in the last term of (116) and
then integrate by parts, so we get a
formula like (70) with
$\frac{d}{dv}(l_n(v) h_{n, t-\tau} (v)) $ instead of $g'$;
$l_n(v) h_{n, t-\tau} (v) $ is
continuous and piecewise $C^1$ and
\be
\mid\frac{d}{dv} ( l_n(v) h_{n, t-\tau} (v)) \mid
\leq L^n l_n^2(v) h_{n, t-\tau} (v)
\en
for $v\neq 0$ (see (31-33)).
However, we get from the integration by parts another term
$$
\int^{t-1}_0 d{\tau}
\int dv \frac{\mid u-v \mid}{2\tau}
R^{\tau}_n (u,v) l_n(v) h_{n, t-\tau} (v)
$$
(previously, this term was cancelled by
the third term in (116),
but here we have taken absolute
values).

Let us bound each term of (116),
by the RHS of (110), first
for $u\geq 0$: for the first term,
it is enough to use (47)
(i.e. $|f'_n(u)|\leq l_n(u)$) and (112).
 For the second term, use (78, 79, 60, 61)
and $\int_0^{t-1}d\tau \chi(\tau \leq L^{-{n\over2}})\leq
 L^{-{n\over2}}$ for the second term of (78).
Then, (110) follows for that term from
\be
L^{n{\cal O}(\delta)} L^{{2n\over3}}\leq
L^{n(1+\delta)} l_n(u)
\en
The third
term in (116) is bounded like the first one,
using the exponential decay in
$\frac{\mid u-v \mid}{\sqrt\tau} $
or in $\frac{\mid u\mid+\mid v
\mid}{\sqrt\tau} $  in (53, 56) and the factor
$\frac{\mid v
\mid}{\sqrt\tau}$ in (56) to
control
$\frac{\mid u-v \mid}{\sqrt\tau}$.
Then we are left with the
integral  $\int d\tau \frac{1}{\sqrt\tau} $,
which is finite. For the terms produced by
the integration by parts, the one with $f'_n(v)$
is controlled using (47), i.e.
$|f'_n(v)|  \leq  l_n(v)$ and
\be
\int^{t-1}_0 d{\tau}  \int dv R_n^\tau (u,v) l_n^2 (v)
h_{n, t-\tau} (v) dv \leq L^{2n
\delta} L^{-\frac{n}{12}}l_n(u) h_{n, t} (u),
\en
which is proven like (112), using, instead of (114),
$$
\int^{t-1}_0 d{\tau}  \int
dv R_n^\tau (u,v) (1+L^n|v|)^{-{1\over 2}}
h_{n, t-\tau} (v) dv \leq L^{n\delta}
 L^{-\frac{n}{3}} h_{n, t} (u)
$$
where the worst term comes from the first term in (58).
To get (119), we used
$ L^{-\frac{n}{3}}\leq L^{-\frac{n}{12}} l_n(u)$.
Still, we have
$L^{-\frac{n}{12}}$ to control
$L^{n {\cal O}(\delta)}$.
The term with $\frac{\mid u-v
\mid}{\tau} $ has already been discussed, while, for
the term with  $\frac{d}{dv}(l_n(v)
h_{n, t-\tau} (v))$, we use (117) and (119).

We still have to prove (110) for $u \leq 0$.
We proceed as in (116), using (82) instead of (115).
All the terms  are bounded as before, except
$$
\int^{t-1}_0 d{\tau}
\int dv \frac{\mid u-v +2 L^n\tau \mid}{2\tau}
R^{\tau}_n (u,v) l_n(v) h_{n, t-\tau} (v) \leq
C h_{n, t} (u) \left( L^{{2n\over 3}}   +
L^{n ({3\over 4}+{\delta\over 4})}l_n(u)
\right)
$$
Indeed, the factor $\frac{\mid u-v +2 L^n\tau \mid}{2\tau}$
can be controlled as before, using the exponential decay
in (57), which gives a bound like (112),
while, in (58), we use
 the factor $L^{-n}\frac{\mid u-v
\mid}{\sqrt\tau}$ and $\int d \tau \frac {1}{\sqrt\tau}\
\chi(\tau \leq L^{-{n\over2}})\leq
  cL^{-{n\over4}}$ .
Then, we use (118) to get (110).

The proof of (107) follows the same
pattern. We want to show: \be
 \mid \int^{t-1}_0 d{\tau} \int dv Q_n^{\tau}
(u,v) N (v,t-{\tau}) \mid \leq  L^{-n\delta} w_{n, t} (u)
\|(\varphi,s) \|^{(n)}
\en
and
\BE
\mid \frac{d}{d u} \int^{t-1}_0 d{\tau}\int dv Q_n^{\tau}
(u,v) N (v,t-{\tau})  \mid   \leq
L^{-n\delta} L^{n(1+\delta)} w_{n, t} (u)
\|(\varphi,s) \|^{(n)}
\EN

Let us prove (120,121) for each term in $N$, see (27).
For $- r_n (\p \varphi_n)^2 $, we use, see (2.5),
\be
r_n \leq c(1+L^nv)(1+e^{L^nv})^{-1}
\en
and $r_n h_n \leq c w_n$, $h_n\leq c L^{-n}$
for $v\geq -\epsilon_n$, combined with  (108), to get
\be
r_n(v)(\p \varphi_n)^2(v, t-\tau)\leq L^{n{\cal O}(\delta)}
w_n(v, t-\tau)l_n(v)\|(\varphi,s) \|^{(n)}
\en
for $v\geq -\epsilon_n$.
Then, we can prove (120), for $u\geq 0$, by using
Lemma 5, a, (102) and
$$
\int^\infty_{-\epsilon_n} dv
D^{\tau}_{-2\epsilon_n} (u,v) (1+v)^{-2 - \delta}
 (1+L^n |v|)^{-{1\over 4}}
\leq C(L) L^{-{n \over 4} }(1+u)^{-2-\delta}
$$
so that we may  use, as before, $L^{-{n \over 4}} $ to control
 $L^{n{\cal O}(\delta)}$.
Now, consider (120) for $u\leq 0$; for the integral over
$v\geq -\epsilon_n$, we use (123) and (101); again,
the integral over $\tau$
gives a small contribution.
For $v\leq -\epsilon_n$, $r_n$ is of order one and
 we have, using (108),
\be
(\p \varphi_n)^2(v, t-\tau) \leq
 L^{n(1+{\cal O}(\delta))}h_{n, t-\tau} (v) l_n(v)
\|(\varphi,s) \|^{(n)}
\en
Now, use (100) to prove:
\be
\int^{t-1}_0 d{\tau}  \int_{-\infty}^{-\epsilon_n} dv
Q_n^\tau (u,v) h_{n, t-\tau} (v) dv \leq
 L^{-n(2-\delta)} h_{n, t} (u)
\en
and
\be
\int^{t-1}_0 d{\tau}  \int_{-\infty}^{-\epsilon_n} dv
Q_n^\tau (u,v) (1+L^n|v|)^{-{1\over 4}}
h_{n, t-\tau} (v) dv \leq
 L^{-n(2+\frac{1}{4}-\delta)} h_{n, t} (u).
\en
These bounds are similar to (113,114) but we use the factor
$e^{-c\tau L^{2n}}$
in (100) and
\be
\int^{t-1}_0  d\tau e^{-c\tau L^{2n}} \leq C L^{-2n}
\en
to get $L^{-2n}$ in (125,126).
Combining (124-126), we get a bound on the integral (120)
 over
$v\leq -\epsilon_n$, for $- r_n (\p \varphi_n)^2 $ and
$u\leq 0$, of the form
$$
L^{-n(1+\frac{1}{4}-{\cal O}(\delta))} h_{n, t} (u)
\|(\varphi,s) \|^{(n)}
$$
Then, (120) is proven for that term
by observing that, for $u\leq 0$,
$L^{-n}h_{n, t} (u) \leq C(L) L^{n\delta}w_{n,t}(u)$.
The proof of (121) combines these bounds
 and the proof of (110) (actually, it is easier, since we do not have
$l_n(u)$ in (121)).
The term $- s_n (\p \varphi_n)^2$ in (27)
 can be treated in the same
way, since $s_n$ is smaller than $r_n$.

The final term to consider is $- 3 r_n L^{2n} s^2_n$,
since the term $-
L^{2n} s^3_n$ is smaller. Following the same strategy
as before, we use, instead of (123, 124), the bound:
\be
r_n(v) L^{2n} s^2_n(v, t-\tau) \leq L^{n{\cal O}(\delta)}
(1+L^nv)^{2}(1+e^{L^nv})^{-2} w_{n, t-\tau}(v)
\|(\varphi,s) \|^{(n)}
\en
which follows from (108, 122) and $w_n\leq L^{-n}$.
Because of the factor $(1+L^nv)^{2}(1+e^{L^nv})^{-2} $, we
get an extra $L^{-n}$ when
we integrate over $v \geq 0$, which is used to
control $L^{n{\cal O}(\delta)}$, and we get $L^{-2n}$
from (127) when we integrate over
$v\leq -\epsilon_n$; the integrals over
$-\epsilon_n \leq v\leq 0$ are
controlled using (98), (101), and
$\int_0^{t-1} d\tau \int^0_{-\epsilon_n} dv
D^\tau_{-2\epsilon_n}(u, v)
\leq c \epsilon_n$.

\vspace*{5mm}

\no{\Large\bf Appendix}

\vspace*{10mm}

\no{\bf Proof of Lemma 3}
\setcounter{equation}{0}
\vspace*{5mm}

To prove (3.53), we use the path space representation (3.51).
For each path $\omega$, let $t_1$ be the time of
the first visit to $-2\epsilon_n$, if
$\omega$ visits $-2\epsilon_n$ before time $\tau$
and let $t_2$ be the time of the
first visit to $- \frac{3}{2} \epsilon_n$ after
$t_1$ (since $v \geq -\epsilon_n,
t_2 \leq \tau$ if $t_1 < \tau$). So, by conditioning,
\BE
&e^{-\tau H_n} (u,v) = \int^\tau_0 dt_1 \int^\tau_{t_1}
dt_2 \int d
\mu^{t_1}_{u,-2\epsilon_n} (\omega>-2\epsilon_n)
d\mu^{t_2-t_1}_{-2\epsilon_n, - \frac{3}{2}\epsilon_n}
 (\omega< - \frac{3}{2}
\epsilon_n)
\non\\
&\cdot d \mu^{\tau-t_2}_{- \frac{3}{2}
\epsilon_n,v} (\omega)
e^{- \int^\tau_0 V_n (\omega(s))ds}
+ \int d \mu^\tau_{u,v} (\omega)
\chi (\omega>-2\epsilon_n)
e^{-\int^\tau_0 V_n (\omega(s))ds}
\EN
where $d \mu^t_{a,b} (\omega>b)$
denotes the measure
on paths with $\omega(0)=a, \omega(t)=b$ and
 $\omega(s)>b$ for $s<t$,  defined by:
$$
\int F (\omega) d \mu^t_{a,b} (\omega > b) =
2 \frac{d}{dx} \int F (\omega) d
\mu^t_{a,x} (\omega) \mid_{x=b}
$$
For our $F$, the RHS is $C^{\infty}$ in $x$ and this indeed
defines an expectation. This equation may be derived by the
method of images from random walks. We will have below
positive $F$'s estimated from above by constants, and
the only thing we need to know is the formula for
$F=1$,
\be
\int d \mu^t_{a,b} (\omega>b) = \frac{(a-b)}
{(4 \pi t^3)^{1/2}} e^{- \frac{(a-b)^2}{4t}}
\en
which is the probability density
that $t$ is the first time at which $\omega$, starting from
$a$, reaches $b$.

Using $V_n \geq 0$, the second term in (1) is
 bounded by $D^\tau_{-2\epsilon_n} (u,v)$
and the formula (3.54) follows from the
method of images. In the
first term, we use
\be
\int^\tau_0 V_n (\omega(s)) ds \geq L^{2n}
(t_2 - t_1) - C
\en
which holds because, for $s \in [t_1 , t_2]$,
 $ \omega(s) < -\frac{3}{2}
\epsilon_n$, $ V(x) \geq 1-Ce^{-c|x|}$,
and $L^{2n} e^{-c\epsilon_n L^n} \leq C/L^2\leq C/\tau$ for
$k$ large in (3.52).
So, using (2), (3), the first term in (1) can be
 bounded by
\be
C \int^\tau_0 dt_1 \int^\tau_{t_1} dt_2
 \frac{(u+2\epsilon_n)}{t^{3/2}_1}
e^{-\frac{(u+2\epsilon_n)^2}{4t_1}}
\frac{\epsilon_n}{(t_2 - t_1)^{3/2}} e^{-
\frac{\epsilon_n^2}{16(t_2-t_1)}}
e^{-L^{2n}(t_2-t_1)} \frac{e^{-\frac{(v+
\frac{3}{2}\epsilon_n)^2}{4(\tau-t_2)}}}{(\tau-t_2)^{1/2}}
\en

Now we use the following three bounds. First,
\be
\frac{e^{-\frac{(v+ \frac{3}{2}\epsilon_n)^2}
{4(\tau-t_2)}}}{(\tau-t_2)^{1/2}}
 \leq \frac{C e^{-c\frac{|v|}{\sqrt {\tau}}}}
{\epsilon_n}
\en
which holds because $v \geq -\epsilon_n$, $\tau-t_2 \leq \tau$. Secondly,
\be
\int^\infty_{0} ds\frac{\mid z \mid}{s^{3/2}}
e^{- \frac{z^2}{4s}}
e^{-\alpha s}\leq C e^{-\sqrt\alpha |z|}
\en
which follows by shifting $s$ by the location of the maximum of
the exponent $\frac{|z|}{2\sqrt\alpha}$. We use (6)
with $z = \epsilon_n$, and
$s = 4(t_2 - t_1)$, $\alpha = \frac{L^{2n}}{4}$. Thirdly,
\be
\int^\tau_0 ds \frac{\mid z \mid}{s^{3/2}}
e^{- \frac{z^2}{4 s}} \leq C e^{-c
\frac{\mid z \mid}{\sqrt{\tau}}} \en
which follows also from
a change of variable, $s'=\frac{s}{z^2}$.
 We use (7) with $z=u+2 \epsilon_n$
and $s=t_1$.

Using (5,6,7),  we get
\be
(4) \leq  \frac{C}{\epsilon_n} e^{-cL^n\epsilon_n}
e^{-\frac{c}{\sqrt{\tau}} (u+v)}
 \leq  L^{-np} e^{- \frac{c}{\sqrt{\tau}} (u+v)}
\en
for $k$ large in (3.52). This completes the proof of (3.53).

To prove (3.56), we use (3.46):
\be
R_n^\tau (u,v) \leq C e^{-f_n (v)} (e^{-\tau  H_n}) (u,v) (1+ L^nu)^{-1}
\en
and, by (3.44, 3.52),
\be
-f_n (v) \leq L^n v +c \leq - \epsilon_n L^n +c =-kn\log L +c
\en
for $v \leq - \epsilon_n$. Then we use $V_n \geq 0$, to bound
\be
e^{-\tau  H_n} (u,v)\leq \int d \mu^\tau _{u,v} (\omega)=
\frac{e^{-\frac{(u-v)^2}{4\tau }}} {\sqrt{4\pi\tau }}
\en
And, since $u\geq 0$, $v\leq -\epsilon_n$, we get
 (3.56) from (9, 10, 11).

To prove (3.57) we use again (3.51).
Let $t_1$ be the time of the first visit of $\omega$
to $- \frac{\epsilon_n}{2}$, if $\omega$ visits
$- \frac{\epsilon_n}{2}$ at all. We have
\BE
R_n^\tau  (u,v)& =&
\exp (f_n (u) - f_n (v))
\left( \int d \mu^\tau _{u,v}(\omega)  e^{-\int^\tau _0 V_n
(\omega(s)) ds} \chi(\omega < -
\frac{\epsilon_n}{2}) \right. \non \\
&+&\left. \int^\tau _0 d t_1 \int d \mu^{t_1}_{u,-
\frac{\epsilon_n}{2}}
(\omega < -\frac{\epsilon_n}{2})
d \mu^{\tau -t_1}_{- \frac{\epsilon_n}{2}, v} (\omega)
e^{- \int^\tau _0 V_n (\omega(s)) ds} \right)
\EN

We use (3.44):
\be
f_n(u) - f_n (v) \leq -L^n (u-v) + c
=L^n(|u|-|v|) +c
\en
for $u,v$ less than zero; for the first term in (12),
 we have, as in  (3),
\be
\int^\tau _0 V_n (\omega(s)) ds \geq \tau  L^{2n} - C
\en
because $\omega(s) < - \frac{\epsilon_n}{2}$.
Using the equality in (11),
the first term of (12) is bounded by the
first one in (3.57).

For the second term in (12), we have
\be
\int^\tau _0 V_n (\omega(s)) ds \geq t_1 L^{2n} - C
\en
and
\be
\int d \mu^{\tau -t_1}_{- \frac{\epsilon_n}{2},v} (\omega)
\leq C (\tau -t_1)^{-\ha}e^{- \frac{c\mid
v \mid}{\sqrt{\tau-t_1} } }\leq \frac{C}{\epsilon_n}
e^{- \frac{c}{\sqrt{\tau} } \mid
v \mid},
\en
which holds because $v \leq -\epsilon_n$.
Using (2) and (15, 16) we bound the second integral in  (12) by
\be
\frac{C}{\epsilon_n}
e^{- \frac{c}{\sqrt{\tau} } \mid
v \mid} \int^\tau_0 dt_1 \frac{\mid u + \frac{\epsilon_n}{2}
\mid}{t_1^{3/2}}
e^{-\frac{(u+ \frac{\epsilon_n}{2})^2}{4t_1} - t_1 L^{2n}}
\en
which, using (6) with $z= u+ \frac{\epsilon_n}{2}$,
 $\alpha = L^{2n}$ is less than
\be
\frac{C}{\epsilon_n} e^{- \frac{c}{\sqrt{\tau} } \mid v \mid}
e^{ - L^n \mid u \mid + L^n \frac{\epsilon_n}{2}}
\en
since $u\leq 0$.
Combining this with (13), and remembering that
$v \leq - \epsilon_n$, i.e.$L^n v + L^n
\frac{\epsilon_n}{2} \leq - L^n \frac{\epsilon_n}{2}$,
the second term of (12) is bounded by
\be
L^{-np} e^{- \frac{c}{\sqrt{\tau} } \mid v \mid}
\en
which is the second term of (3.57) for $x>0$. For $x<0$,
we have, see (3.21-3.23)
$\mid u \mid = 2 (\tau +1) L^n + \mid x \mid$
and the
factor $\frac{(u+ \frac{\epsilon_n}{2})^2}{4t_1} + t_1 L^{2n}$
 in (17)
reaches its minimum  at $ t_1 =
\tau$, where
\be
\frac{(u+ \frac{\epsilon_n}{2})^2}{4\tau} + \tau L^{2n}
\geq L^n \mid u \mid - \frac{L^n \epsilon_n}{2} +
 c\frac{\mid x \mid}{\sqrt{\tau}}
\en
which then gives (3.57).

Finally, we prove (3.58). Let $t_1$ be
 the time of the first visit of $\omega$
to zero. The LHS of (3.58) is bounded by
\BE
&C& L^{-n(1+\delta)}e^{L^n|u|}
\left( \int^\infty_{-\epsilon_n} dv
(1+L^{-n}
\frac{|u-v|}{\sqrt{\tau}})
\int^\tau_0
dt_1 \int d \mu^{t_1}_{u,0} (\omega<0) e^{-
\int^\tau_0 V_n (\omega(s)) ds}
d \mu^{\tau-t_1}_{0,v} (\omega) \right. \non \\
&+&\left. \int^0_{-\epsilon_n} dv e^{-L^n|v|}
(1+L^{-n}
\frac{|u-v|}{\sqrt{\tau}})
\int
d \mu^{\tau}_{u,v} (\omega)
\chi (\omega<0) e^{- \int^\tau_0 V_n
(\omega(s)) ds}  \right)
\EN
where the second term collects walks that
do not visit zero, in which case $v$ has to be
negative. We used :
$$
e^{-f_n(v)} h_n (v) \leq C L^{-n(1+ \delta)}
$$
for $v \geq - \epsilon_n$ (see (3.61)), in the first term and
$$
e^{-f_n(v)} h_n (v) \leq C L^{-n(1+ \delta)} e^{-L^n |v|}
$$
for $-\epsilon_n \leq v \leq 0$ in the second term.

We shall bound each term of (21) by
\be
C(L) L^{-n(1+\delta)} (L^{-n (\frac{1}{3} + \delta)} +
 L^{\frac{n \delta}{8}} l_n (u) \chi (\tau \leq
L^{-n/2}) ) e^{-\frac{c}{\sqrt \tau} |x| \chi (x <0)}
\en
and, to get (3.58), use
$$
L^{-n(1+\delta)} e^{-\frac{c}{2\sqrt \tau} |x| \chi (x<0)}
 \leq C(L) h_{n, t} (u)
$$
for $u<0$, which holds since
$x \leq 2 t L^n$ for $ x>0$, $u < 0$, and then, use (3.43).

For the first term of (21), we write
\be
\int^\tau_0 V_n = \int^{t_1}_0 V_n +
\int^\tau_{t_1} V_n
\en
and we have, by (3.53),
\be
\int d \mu^{\tau-t_1}_{0,v} (\omega)
e^{- \int^\tau_{t_1} V_n} \leq
D^{\tau-t_1}_{-2\epsilon_n} (0,v) +
L^{-np} e^{- c \frac{\mid v \mid}{\sqrt\tau}}
\en
Besides, from the explicit formula (3.54), one gets

\be
\int^\infty_{-\epsilon_n} dv
D^{\tau-t_1}_{-2\epsilon_n} (0,v)(1+\frac{|v|}{\sqrt{\tau}})
 \leq C \min \left(
\frac{\epsilon_n}{{\sqrt{\tau-t_1}}},1 \right)
\en
and, after integration over $v$,
the second term in (24) is smaller.
Also, we shall prove below that

\BE
\int d \mu^{t_1}_{u,0} (\omega<0) e^{- \int^{t_1}_0 V_n}
\leq
\left\{
\begin{array}{lll}
C \frac{|u|}{t_1^{3/2}} e^{- \frac{u^2}{4t_1}
 -L^n |u|-ct_1 L^{2n}} & \mbox{if} & t_1 >
\frac{4 |u|}{L^n} \\
C \frac{|u|}{t_1^{3/2}} e^{- \frac{u^2}{4t_1} -
t_1 L^{2n}} & \mbox{if} &t_1 \leq
\frac{4 |u|}{L^n}
\end{array}
\right.
\EN

Let us consider $\tau \leq \frac{4 |u|}{L^n}$.
Using (24-26), we bound the first integral in (21) by
\be
C (1+ L^{-n}\frac{|u|}{\sqrt\tau})
\int^\tau_0 dt_1 \frac{|u|}{t^{3/2}_1}
e^{- \frac{u^2}{4t_1} - t_1 L^{2n}}
\min \left(  \frac{\epsilon_n}{\sqrt
{\tau-t_1}} , 1 \right)
\en

Consider first $|u| \leq L^{-n}$. Then we bound
the minimum in (27)  by $1$,
the integral is bounded by
$e^{-c\frac{|u|}{\sqrt\tau}}$, using (7),
and this is all we need in
(22), since here $x>0$, $\tau \leq 4 L^{-2n} \leq
L^{- \frac{n}{2}}$ and $l_n (u) \geq
c$ for $|u| \leq L^{-n}$. So, let  $|u| \geq
L^{-n}$, and bound the minimum in
(27) by $\frac{\epsilon_n}{\sqrt{\tau-t_1}}$.
Then, change variables :
\BE
u = L^{-n} \tilde{u},\;\;
t_1 = \frac{|u|s}{2L^n} = \frac{|\tilde u| s}
{2 L^{2n}} \;\;
\tau = \frac{|u| \sigma}{2 L^n}  =
\frac{| \tilde u| \sigma}{2 L^{2n}} \nonumber
\EN
Note, that now $ | \tilde u| \geq 1$. We get the integral
\be
\frac{\epsilon_n L^n}{\sqrt{| \tilde u |}}
(1+ cL^{-n}\frac{\sqrt {| \tilde u |}}{\sqrt \sigma})
\int^\sigma_0  ds \frac{\sqrt {| \tilde u |} e^{-
\frac{| \tilde u |}{2} (s+ \frac{1}{s})}}
{s^{3/2} \sqrt {\sigma -s}}
\en
which we want to bound, see (21,22), for $x \geq 0$, by

\be
e^{-| \tilde u |} L^{-n(\frac{1}{3} + \delta)}
\hspace{15mm} \mbox{for} \;\;\; \sigma \geq \frac{2
L^{\frac{3n}{2}}}{| \tilde u |}
\en
and by
\be
\frac{L^{\frac{n \delta}{8}} e^{- | \tilde u |}}
{(1+ | \tilde u |)^{1/4}}
\hspace{15mm} \mbox{for} \;\;\; \sigma \leq \frac{2
L^{\frac{3n}{2}}}{| \tilde u |}
\en
We divide the integral in (28) into $| s-1 | \leq
\frac{1}{4}$ and $| s-1| \geq \frac{1}{4}$. For
the first part, a Gaussian integral around $s=1$ gives
\BE
\int^{ \frac{5}{4}}_{\frac{3}{4}} ds
\frac{\sqrt {| \tilde u|} e^{- \frac{| \tilde u|}{2}
(s+ \frac{1}{s})}}{\sqrt {\sigma-s}} \leq
 c | \tilde u |^{1/4} e^{- | \tilde u |}
\EN
and, if $\sigma \geq \frac{3}{4}$, $\tau \leq L^2$
implies that $| \tilde u|\leq C(L) L^{2n}$, so that
$(1+ cL^{-n}\frac{\sqrt {| \tilde u |}}{\sqrt \sigma})
\leq C(L)$,
while
\BE
\int^\sigma_0 ds
\frac{\sqrt {| \tilde u|} e^{- \frac{| \tilde u|}{2}
(s+ \frac{1}{s})}}{s^{3/2} \sqrt {\sigma-s}}
\chi (| s-1 | \geq \frac{1}{4}) \leq c e^{- | \tilde u
|}
\EN
where we use $s+ \frac{1}{s} \geq 2 + \frac{c}{s}$
for $| s-1 | \geq \frac{1}{4}$ and we use the
factor $\exp (-\frac{c | \tilde u |}{2s})$ to control
 $\frac{\sqrt {| \tilde u |}}{s^{3/2}}$
or $\frac{\sqrt {| \tilde u |}}{\sqrt\sigma}$
(since $| \tilde u | \geq 1$).

Using (31,32), we get that (28) is bounded by
$$
C(L) \epsilon_n L^n | \tilde u |^{- \frac{1}{4}}
e^{-| \tilde u |}
$$
which is always less than (30) since $| \tilde u | \geq 1$.
And, for $\sigma \geq \frac{2 L^{\frac{3n}{2}}}
{| \tilde u |}$, we have $| \tilde u | \geq c L^{
\frac{3n}{2}}$, since $\tau \leq \frac{4 | u |}{L^n}$
means $\sigma \leq 8$. So, $| \tilde u |^{-
\frac{1}{4}} \leq L^{-\frac{3n}{8}}$ and, inserting
this in (33), gives (29), for $\delta$ small.

For $x \leq 0$, we have $| u | \geq 2 (\tau + 1) L^n, \tau
\leq L^2-1$, so that $\sigma \leq 1 -
\frac{c}{L^2}$. Since $\sigma  $ is bounded away from
 one, we can improve the bound in (31,32) into $c
e^{-| \tilde u | (1+ \frac{c}{L^2})}$. But
$\frac{| \tilde u |} {L^2} \geq |x|$ for $x <0$, so
we get the exponential decay needed in (22).

For $\tau \geq \frac{4 |u|}{L^n}$, we get from (24-26)
 and from $L^{-n}\frac{\sqrt {|  u |}}{\sqrt \tau}
\leq {\sqrt \tau}
\leq L$
an upper bound on the first integral in
(21) of the form :
\BE
& &C(L)
\left(
\int^{4 L^{-n} |u|}_0 dt_1
\frac{|u|}{t^{3/2}_1}
e^{- \frac{u^2}{4t_1} - t_1 L^{2n}}
\min \left(  \frac{\epsilon_n}{\sqrt
{\tau-t_1}} , 1 \right)\right. \non \\
&+& \left. e^{-L^n |u|}
\int^\tau_{4 L^{-n} |u|} dt_1 \frac{|u|}{t^{3/2}_1}
e^{-\frac{u^2}{4t_1} - ct_1 L^{2n}}
\min \left(  \frac{\epsilon_n}{\sqrt
{\tau-t_1}} , 1 \right)
\right)
\EN

The first term can be analyzed as we did
with (27) and the second term is always bounded by $ e^{-
L^n |u|} e^{-c L^n |u|}$ and is bounded by $c\epsilon_n L^{n\over4}e^{-
L^n |u|} e^{-c L^n |u|}$ for $\tau\geq L^{-n\over2}$. These bounds are
less than (29, 30).

For the second integral in (21), we introduce
$1=\chi (v > u) + \chi (v \leq u)$ and, for $v >
u$, we let $t_1$ be the time of the first visit to $v$.
Then, we can bound that integral by
\begin{eqnarray}
& &\int^0_{-\epsilon_n} dv \chi (v > u) e^{- L^n | v|}
 (1+ L^{-n}\frac{|u-v|}{\sqrt \tau})
\int^\tau_0 dt_1 d \mu^{t_1}_{u, v} (\omega < v)
 e^{- \int^\tau_0 V_n (\omega (s) ds} d
\mu^{\tau -t_1}_{v,v} (\omega) \nonumber\\
&+& c \int^0_{- \epsilon_n} d v \chi (v \leq u) e^{-L^n |v|}
(1+ L^{-n}\frac{|u-v|}{\sqrt \tau})
\frac{e^{-\frac{(u-v)^2}{4 \tau}}}{\sqrt \tau}
\end{eqnarray}

The second term is bounded by
$$
c \min (1, \frac{L^{-n}}{\sqrt \tau}) e^{-L^n |u|}
$$
which gives a contribution to (22) since, for
$|u| \leq  \epsilon_n$, $l_n(u)\geq (1 + L^n |u|)^{-
\frac{1}{4}} \geq L^{- \frac{n \delta}{8}}$.
For the first term in (34), we use:

\BE
\int d \mu^{t_1}_{u,v}  (\omega<v)
e^{- \int^\tau_0 V_n (\omega(s))ds}
\leq
\left\{
\begin{array}{ll}
C \frac{|u - v|}{t_1^{3/2}} \exp
 \left( - \frac{(u-v)^2}{4 t_1} -
 L^n (|u -v|)-ct_1 L^{2n}\right) &
 \mbox{if} \; t_1 >
\frac{4 |u - v|}{L^n} \\
C \frac{|u - v|}{t_1^{3/2}} \exp
 \left( - \frac{(u-v)^2}{4 t_1} - t_1 L^{2n} \right) &
\mbox{if} \; t_1 \leq \frac{4 |u-v|}{L^n}
\end{array}
\right.
\EN
and argue
as for the first integral in (21).

So, we are left with the proof of (26).
The proof of (35) is similar.
Consider first $t_1 \geq \frac{4|u|}{L^n}$.
Using (3.44) we have $r^2_0 (u) \geq 1-c
\lambda \geq 1/2$ for $u<0$, so
\be
\int^{t_1}_0 V_n (\omega (s)) ds \geq
\frac{L^{2n}t_1}{2} \geq  L^n |u| +c t_1 L^{2n}
\en
which, combined with (2),
gives (26) for that case.

For $t_1 \leq \frac{4 |u|}{L^n}$, we write
\be
\int^{t_1}_0 V_n (\omega (s)) ds \geq L^{2n}t_1 -
 \lambda \int W_n (\omega (s)) ds
\en
where
\be
|W_n (\omega (s)) | \leq L^{2n} e^{-c L^n | \omega (s) |}
\en
and, as in (3.44), the origin of the coordinates
is chosen so that
$\lambda$ is small enough. So to prove
(26), we have to show (see (2))
\be
\langle e^{\lambda \int^{t_1}_0 W_n (\omega (t)) ds}
\rangle \leq C
\en
where the expectation value and the probability
 $P$ below refer to the measure
$d \mu^{t_1}_{u,o} (\omega < 0)$.
We bound the LHS of (39) by
\be
\sum_{m \geq 0} e^{\lambda (m+1)}
P (| \int^{t_1}_0 W_n (\omega (s)) ds | \in [m, m+1])
\en
Define $t_k (\omega) = |\{s |\omega (s) \in
[\frac{-k}{L^n} , \frac{-k+1}{L^n}]\}|$ for
$k\geq1$.
By (38), $\int^{t_1}_0 W_n (\omega (s)) ds \leq L^{2n}
 \sum_k t_k e^{-c(k-1)}$; so, we can
bound (40) by
$$
\sum_{m \geq 0} e^{\lambda (m+1)}
\sum_{k\geq1} P (t_k(\omega)
\geq \alpha m k L^{-2n})
$$
where $\alpha$ is such that $\alpha
\sum_k k e^{-c(k-1)} \leq 1$.
We shall prove that
\be
P (t_k (\omega)\geq \alpha m k L^{-2n})
\leq Ce^{-c \alpha m k}
\en
for $m$ large. This then proves (39) for $\lambda$ small
enough.

Write
\be
\int d \mu^{t_1}_{u,0} (\omega < 0) = \int_0^{t_1} ds \int
d \mu^{t_1 -s}_{u,\frac{-k}{L^n}}
(\omega < - \frac{k}{L^n}) d \mu^s_{-\frac{k}{L^n},0}
 (\omega <0)
\en
i.e. $t_1 - s$ is the time of the first visit of
 $\omega$ to $\frac{-k}{L^n}$.
Clearly,
$$
\chi(t_k(\omega)\geq \alpha m k L^{-2n}) \leq
 \chi (s (\omega) \geq \alpha m k
L^{-2n})
$$
So that to prove (41) we will show (see (2))
\be
\int^{t_1}_{\alpha mkL^{-2n}} ds \int
d \mu^{t_1 -s}_{u,\frac{-k}{L^n}}
(\omega < - \frac{k}{L^n}) d \mu^s_{-\frac{k}{L^n},0}
(\omega <0) \leq C e^{-c \alpha
mk} \frac{|u|}{t_1^{3/2}} e^{- \frac{u^2}{4 t_1}}
\en
where the LHS is
\be
C \int^{t_1}_{\alpha mkL^{-2n}} ds
\frac{|u+ \frac{k}{L^n} |}{(t_1-s)^{3/2}}
e^{-\frac{(u+\frac{k}{L^n})^2}{4(t_1-s)}}
\frac{k}{L^n s^{3/2}}
e^{-(\frac{k}{L^n})^2/4s}
\en
Consider (43) for $u\leq-\frac{2k}{L^n}$,
otherwise $t_1\leq \frac{4|u|}{L^n}
\leq \frac{8k}{L^{2n}}\leq \frac{\alpha m k}{L^{2n}}$
for $m$ large, and (43) is trivial.
Now use $t_1-s = t_1 (1-
\frac{s}{t_1})$,  $\frac{1}{1-\frac{s}{t_1}} \geq 1+
\frac{s}{t_1}$, $1+ \frac{s}{t_1} \leq 2$, to get
\be
\frac{(u+ \frac{k}{L^n})^2}{4 t_1 (1- \frac{s}{t_1})}
\geq \frac{u^2}{4 t_1} +
\frac{s}{4 t^2_1} u^2 - \frac{|u| k}{t_1 L^n}
\en
which, for $t_1 \leq \frac{4 |u|}{L^n}$ and $s
\geq \alpha m k L^{-2n}$, i.e. for $\frac{s |u|}{t_1}\geq
 \frac{\alpha m k L^{-n}}{4}$,
 is, for $m$ large, bigger than
\be
\frac{u^2}{4t_1} + \frac{s}{8t^2_1} u^2 \geq
\frac{u^2}{4t_1} + c \alpha m k
\en
Now consider separately the integral in (44) over
$s$ so that $t_1-s \geq \eta t_1$ and
$t_1-s \leq \eta t_1$, where $\eta$ is small,
say less than $\frac{1}{16}$.
 For $t_1-s \geq \eta t_1$ we use $|u+
\frac{k}{L^n} | \leq |u|$ (since $u \leq -\frac{2k}{L^n}$) and
(46) to bound that integral by
$$
C \frac{|u|}{t^{3/2}_1} e^{-\frac{u^2}{4t_1}}
 e^{-c \alpha m k} \int ds
\frac{k}{L^n s^{3/2}} e^{-(\frac{k}{L^n})^2/4s}
$$
where the last integral is of order one by (7).
So we get a contribution to (43). For
$t_1-s \leq \eta t_1$, with $\eta$ small, we have
for $u\leq -\frac{2k}{L^n}$, $|u+\frac{k}{L^n}|\geq \ha|u|$,
and
$$
\frac{(u+\frac{k}{L^n})^2}{4(t_1-s)} \geq
\frac{(u+ \frac{k}{L^n})^2}{8 (t_1-s)} +
\frac{u^2}{2t_1};
$$
 since $s \leq t_1$, $\frac{u^2}{4t_1}
\geq \frac{su^2}{4t^2_1} \geq c \alpha
mk$
where the second inequality comes from
$t_1 \leq \frac{4 |u|}{L^n}$, $s \geq \alpha m k L^{-2n}
$.
So  we may bound the integral (44) over
$t_1-s \leq \eta t_1$, by
\be
C e^{-\frac{u^2}{4t_1}} e^{-c \alpha mk} \int_0^{t_1} ds
\frac{|u+ \frac{k}{L^n}|}{(t_1-s)^{3/2}}
e^{-\frac{(u+ \frac{k}{L^n})^2}{8(t_1-s)}}
\frac{k}{L^n s^{3/2}} e^{-(\frac{k}{L^n})^2/8 s}
\en
but, going back to (42) and using (2),
we see that the last integral
is proportional to
\be
\frac{|u|}{t^{3/2}_1} e^{-\frac{u^2}{8t_1}}
\leq \frac{|u|}{t^{3/2}_1}
\en
(we apply (42) to a Wiener measure with a
diffusion constant whose value is twice the original one); (48)
inserted in (47) gives (43).

\vspace*{5mm}

\no{\bf Proof of Lemma 4}

\vspace{5mm}

To prove (3.78) consider first
$u \geq 2 \epsilon_n$ and let $t_1$
 be the time of the first visit to
$\epsilon_n$.
We write
$$
k_n (u,v,\tau) \leq c L^n \int^\tau_0 dt_1 \int
 d \mu^{t_1}_{u,\epsilon_n} (\omega > \epsilon_n) d
\mu^{\tau-t_1}_{\epsilon_n, v} (\omega)
e^{- \frac{1}{2} \int^\tau_{t_1} V_n (\omega (s)) ds}
$$
\be
+ L^{3n} \int d \mu^\tau_{u,v} (\omega) \chi (\omega >
 \epsilon_n) \int^{\tau}_0 |V' (L^n \omega
(s)) | ds
\en
where we use (3.105) and $V \geq 0$.

We can see that (24,25) hold with 0 replaced by
$\epsilon_n$ and $V$ by $\frac{1}{2} V$ so that
the first term of (49) gives a bound on
$\int^\infty_{-\epsilon_n} dv
 k_n (u,v,\tau)$ of the form
$$
C L^n \int^\tau_0 dt_1 \frac{|u-\epsilon_n|}{t^{3/2}_1}
e^{- \frac{(u-\epsilon_n)^2}{4t_1}} \min
(\frac{\epsilon_n}{\sqrt {\tau-t_1}} , 1)
$$
 where we used also (2). This integral is itself bounded by
$$
\frac{L^{\frac{n \delta}{4}}}{\sqrt \tau}
 e^{- \frac{c | u - \epsilon_n|}{\sqrt \tau}}
$$
which gives a contribution to the RHS of (3.78):
for $\tau \leq L^{-n\over2}$, we use
$
\frac{1}{\sqrt \tau} \leq \frac{C}{|u-\epsilon_n|}
e^{\frac{c}{2} \frac{|u-\epsilon_n|}{\sqrt
\tau}}$ and $|u-\epsilon_n|^{-1}
\leq L^n l_n(u)$ for $u \geq 2 \epsilon_n$.

For the second term in (49), we have
$$
\int^\tau_0 |V' ( L^n \omega (s)| ds \leq L^{-np}
$$
since $\omega \geq \epsilon_n$.
To obtain the factor $e^{- \frac{c}{L} u}$ in (3.78),
let $x=x (\omega)$ be the point closest to
zero which is visited by $\omega$. Then we have
$$
\int^\tau_0| V' (L^n \omega (s))| ds \leq L^{-np} e^{-cL^n x}
$$
and we get a decay $e^{- \frac{c}{\sqrt \tau} |u-x|}$
from the Wiener measure,
for the paths going in time less than
$\tau$ from $u$ to $x$. Finally, for $u \leq 2
\epsilon_n$, we use (3.105) and (24, 25)  with
$0$ replaced by $u$ and $V$ replaced by $\ha V$,
to get
$$
\int dv k_n (u,v,\tau) \leq c L^n \min
(\frac{\epsilon_n}{\sqrt \tau},1)
$$
which again gives a contribution to (3.78)
$(l_n (u) \geq (c(L) n)^{-1/4}$ for $u \leq 2
\epsilon_n$, so we take  $1$ as an upper bound for
$\tau\leq L^{-n\over2}$ and
$\frac{\epsilon_n}{\sqrt \tau}$ otherwise).

The proof of (3.79) is similar to the one
of (3.56), using (3.105).
The proof of (3.80) follows the one of (3.58).
To bound $L^{3n} \int^\tau_0 | V' (L^n \omega
(s)) | ds$, we use $L^{2n} | V' (L^n \omega (s))
| \leq W_n (\omega (s))$ for $\omega (s)
<0$ where $W_n$ occurs in (37,38). Indeed, the only
property of $W_n$ that we used was (38), which
holds for $V'$. Then, we use
$$
L^{3n} \int_0^\tau | V' (L^n\omega (s)) | ds \leq
 L^n c (\lambda) e^{\frac{\lambda}{2} \int^{t_1}_0
W_n (\omega (s)) ds} e^{\frac{1}{2}
\int^\tau_{t_1} V_n (\omega (s)) ds}
$$
where $t_1$ is the first time that $\omega$
visits 0, as in (21), and we use (3.105) for $s \geq
t_1$. Then we can repeat the proof of (3.58)
 and use $c (\lambda) L^n \leq L^{n(1+
\frac{\delta}{4})}$ to obtain (3.80).

The proof of (3.81) follows the one of (3.57) :
consider (12); when $\omega < -
\frac{\epsilon_n}{2}$, we have
$$
\int^\tau_0 | V' (L^n \omega (s)) | ds \leq L^{-np}
$$
and we can get the exponential decay as in the
proof of (3.78). On the other hand, the second
term in (12) led to the second term in (3.57)
which, after integration over $v$ contributes to
the RHS of (3.81).

\vspace*{5mm}

\no{\bf Proof of Lemma 5}

\vspace{5mm}

Part a is proven just like in Lemma 3, with the only
change that $f_n (u)$ is replaced by $-L^n
u$.

The proof of (3.100) is similar to the one of (3.57),
but for $\omega < -
\frac{\epsilon_n}{2}$, we have of course
$$
3 \int^\tau_0 V_n (\omega (s)) ds \geq 3 \tau L^{2n} - c
$$
instead of (14). So we can take $c$ in (3.100)
equal to 2 or to $\alpha-1$, if 3 in (3.25) is
replaced by $\alpha$.

Finally, for (3.101) we can follow the proof of
 (3.58).

\vs{5mm}

\no{\bf Acknowledgments}. During the course of this work, J. B.
has benefitted from the hospitality of the Mathematics Dept of
Rutgers Univrsity, and of the University of Helsinki.
 A. K. benefitted from the hospitality of the University of Louvain.
This work was supported, in part, by the EC Grant SC1-CT91-0695 and the
NSF Grant DMS-8903041.

\vs{10mm}

\end{document}